\documentclass[a4paper,12pt]{article}
\usepackage{epsfig}
\usepackage{amssymb}
\usepackage{amsfonts}
\usepackage{subfigure}
\usepackage{float}
\usepackage{latexsym,hyperref}

\title{Fixed twist dynamics of $SO(3)$ Gauge Theory}
\author{A.~Barresi\thanks{IDS Laboratories S. Piero a Grado (Italy)}\ and
G.~Burgio\thanks{Institut f\"ur Theoretische Physik 
Universit\"at T\"ubingen (Germany)}}

\begin{document}

\maketitle

\begin{abstract}
We perform a throughout study of 3+1 dim. $SO(3)$ LGT for
fixed-twist background. We concentrate in particular on the physically
significant trivial and 1-twist sectors. Introducing a $\mathbb{Z}_2$
monopole chemical potential the 1$^{\rm st}$ order
bulk transition is moved down in the strong coupling region and weakened to
2$^{\rm nd}$ order in the 4-dim
Ising model universality class. In this extended phase diagram we gain access
to a confined phase in every fixed twist sector of the theory.
The Pisa disorder operator is employed together with the Polyakov loop to
study the confinement-deconfinement transition in each sector. Due to the 
specific properties of both operators, most results can be used to gain 
insight in the ergodic theory, where all twist sectors should be summed upon. 
An explicit
mapping of each fixed twist theory to effective positive plaquette models 
with fixed twisted boundary conditions is applied to better establish their 
properties in the different phases.
\end{abstract}

\section{Introduction}
%---------------------
It is common knowledge from lattice investigations that Yang-Mills theories 
possess a finite temperature transition from a confined to a deconfined phase 
\cite{McLerran:1981pk,Kuti:1981gh} linked to the spontaneous breaking of 
center symmetry \cite{Polyakov:1978vu,Susskind:1979up}. For $SU(2)$ 
such transition is 2$^{\rm nd}$ order and therefore in the universality class 
of Ising 3-d \cite{Svetitsky:1982gs}. The
alleged preferential r\^ole that the discretization in the fundamental 
representation plays in such result has been widely discussed in the
literature (see e.g. \cite{Smilga:1993vb}). 

The difficulties connected to
the use of the adjoint discretization, which according to universality
should deliver the same results as the fundamental one for the observables 
which have a common representation,
have been widely discussed and partially understood for a long time 
\cite{Bhanot:1981eb,Greensite:1981hw,Halliday:1981te,Halliday:1981tm}:
the theory exhibits a bulk transition along the adjoint coupling
linked to the condensation of $\mathbb{Z}_2$ magnetic mo\-no\-po\-les, whose 
Dirac strings correspond to open $\mathbb{Z}_2$ magnetic vortices; 
at the same time it was pointed out how the introduction of ad-hoc chemical 
potentials could affect the phase diagram and give access to the continuum 
limit in the weak coupling phase 
\cite{Halliday:1981te,Halliday:1981tm,Caneschi:1982ik}.
Interestingly enough, such topological defects proved also to be the key to 
understand a further property of the adjoint discretization: in the phase 
where $\mathbb{Z}_2$ monopoles condense the $SO(3)$ partition function
with periodic boundary conditions (b.c.) should be equivalent to 
the sum of $SU(2)$ partition functions with all possible 
twisted b.c. \cite{'tHooft:1979uj,Mack:1979gb,Tomboulis:1980vt}.
In the center blind adjoint discretization maximal 't~Hooft loops are
therefore physical topological excitations rather than boundary constraints 
as in
the fundamental one \cite{'tHooft:1979uj,Tomboulis:1980vt,deForcrand:2002vs}.
First attempts to simulate the modified pure adjoint theory proposed in 
\cite{Halliday:1981te,Halliday:1981tm,Caneschi:1982ik} were performed in 
\cite{Cheluvaraja:1996zn,Datta:1996pi,Datta:1997nv,Datta:1999ep,Datta:1999np}. 
Due to the absence of a suitable order parameter the authors had to rely on 
thermodynamic quantities, making the study of the finite
temperature phase transition and its continuum limit quite demanding
\cite{Datta:1999np}. Moreover, it was observed that for small
chemical potential and on top of the bulk transition the theory exhibits 
states where the adjoint Polyakov loop 
$L_A = -1/3$ \cite{Cheluvaraja:1996zn,Datta:1997nv}. In 
\cite{deForcrand:2002vs} it was pointed out how such phase actually 
corresponds to
the non trivial twist sectors of the theory upon which the partition
function should be summed, the analysis in Ref.~\cite{Datta:1999np} neglecting
such aspect. High barriers in the weak 
coupling phase among such 
different topological sectors make an ergodic sampling of the partition 
function very 
difficult already for small volumes \cite{deForcrand:2002vs}, leaving the 
problem of the behaviour of the full adjoint theory open. It was
only recently that consistent efforts through parallel tempering have led to
first results in the ergodic theory \cite{Burgio:2005xe,Burgio:2006dc,Burgio:2006xj}.

In a series of papers 
\cite{Barresi:2001dt,Barresi:2002un,Barresi:2003jq,Barresi:2003yb,Barresi:2004qa,Barresi:2004gk} the analysis of the trivial twist sector was 
developed by studying the spatial distribution of the fundamental 
Polyakov loop $L_F$ \cite{Barresi:2003jq} and the Pisa disorder
operator \cite{Barresi:2004qa}. 
In this paper we will refine and extend such results to the $SO(3)$ dynamics 
of both trivial and non-trivial twist sectors. In particular, in 
\cite{deForcrand:2002vs} it was argued that any configuration generated 
by an adjoint weight at fixed twist could be ``gauge fixed''  
to a configuration kinematically equivalent to a fundamental positive 
plaquette model \cite{Fingberg:1995ut} 
with corresponding twisted b.c.; whether a corresponding effective action can be 
written is however an open question. In Sec.~\ref{sec4} for each fixed 
twist sector an explicit mapping to such positive plaquette model 
configurations will be given. This enables us to define a non-vanishing
$L_F$ and determine the properties of the deconfinement 
phase transition at fixed twist with ``standard'' methods. The intrinsic 
limitations of such procedure will be also discussed. 
The interest of our extensive fixed twist analysis will be made clear in 
Sec.~\ref{sec3}: fixed twist results can deliver 
``low cost'' informations about the ergodic behaviour of some observables in 
regimes hard to investigate with the full $SO(3)$ partition function 
\cite{Burgio:2006dc,Burgio:2006xj}.

\section{Action and observables}

\begin{figure}[thb]
\begin{center}
\subfigure[]{
\includegraphics[angle=0,width=0.8\textwidth]{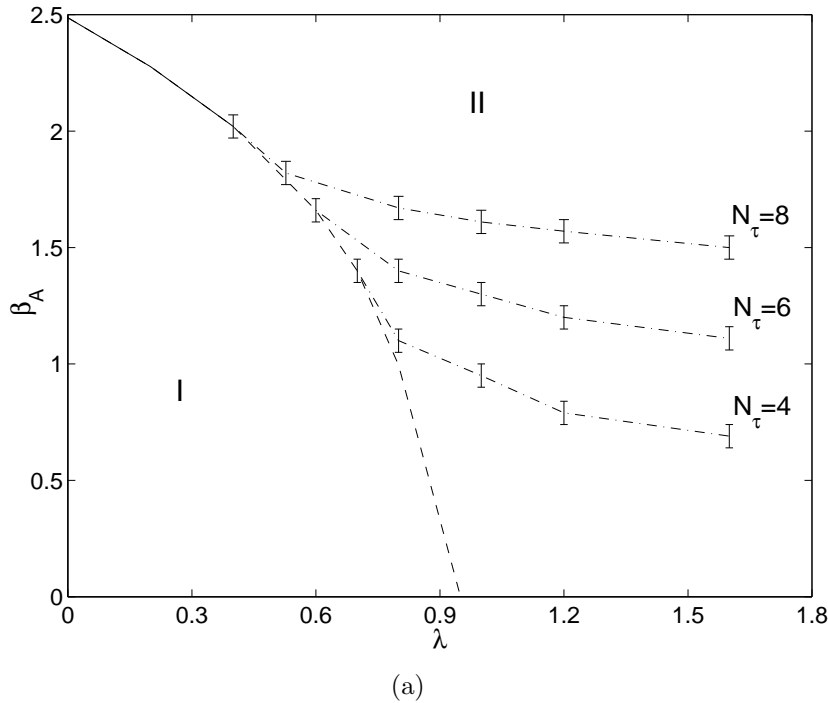}}
\end{center}
\caption{$\beta_A-\lambda$ plane. The bulk transition is shown together with
the finite temperature lines as found at trivial twist in 
\cite{Barresi:2003jq,Barresi:2004qa}}
%\vspace{1cm}
\label{plane}
\end{figure}

We shall study the adjoint Wilson action modified by a $\mathbb{Z}_2$ 
monopole suppression term:
\begin{eqnarray} 
S=\beta_{A} \sum_{P} 
  \left(1-\frac{\mathrm{Tr}_{A}U_{P}}{3}\right)
  +\lambda \sum_{c}(1-\sigma_{c})\,,
\label{ouraction}
\end{eqnarray}
where $U_P$ denotes the standard plaquette and 
$\mathrm{Tr}_{A} = 2 \mathrm{Tr}_{F}^2$ - 1. The product
$\sigma_{c}=\prod_{P\in\partial c}\mathrm{sign}(\mathrm{Tr}_{F}U_{P})$
taken around all $N_c$ elementary 3-cubes $c$ defines the 
$\mathbb{Z}_2$ magnetic char\-ges. Action~(\ref{ouraction}) is center-blind
in the entire $\beta_A-\lambda$ plane (Fig.~\ref{plane}) \cite{Barresi:2003jq}. 
The density 
$M = 1-\langle\frac{1}{N_c}\sum_{c}\sigma_{c}\rangle$
tends to one in the strong coupling region (phase I)
and to zero in the weak coupling limit (phase II). In the $(\lambda,\beta_A)$ 
plane such phases are separated
by a bulk phase transition whose order weakens from the strong 1$^{\rm st}$ 
order at $\lambda=0$, $\beta_A \simeq 2.5$ to 
2$^{\rm nd}$ as $\lambda$ increases 
\cite{Halliday:1981te,Datta:1999np,Barresi:2003jq}. 
For low $\beta_A$ the theory can be shown to be dual to a 4-d Ising model
\cite{Halliday:1981te}, the bulk line terminating at $\lambda \simeq 0.92$.

As already an\-ti\-ci\-pa\-ted, maximal 't~Hooft loops can be generated dynamically in
such adjoint theory, since periodic b.c. on the adjoint fields 
automatically include all twisted b.c. for their fundamental representatives.
Appropriate twist observables can be introduced through
\begin{eqnarray}
z_{\mu\nu}\equiv\sum_{\rho\sigma}\frac{1}{N_\rho N_\sigma}
                \prod_{P\;\in\; \mathrm{plane}\; \mu\nu}
\mathrm{sign Tr}_F U_P\,,
\hspace{.1cm}(\epsilon_{\rho\sigma\mu\nu}=1)\,.
\label{deftwi}
\end{eqnarray}
Like the $\mathbb{Z}_2$ monopoles, such observables are center blind 
\cite{deForcrand:2002vs,Barresi:2003jq}.
In Ref.~\cite{Mack:1979gb,Tomboulis:1980vt} it was shown how the constraint 
$\sigma_{c}=1$, identically satisfied in phase II, assures that $z_{\mu\nu}$ in
eq.~(\ref{deftwi}) can only take the values $\pm 1$; moreover the partition
function generated by action~(\ref{ouraction}) should be equivalent to the 
sum of all partition functions generated with fundamental action and twisted b.c.
Since at finite temperature the 
production of non trivial space like twists ($\mu$, $\nu \neq 4$) will be 
exponentially suppressed, we can concentrate on time like twists 
$z_{i4}$, ($i=1$, 2, 3). We will 
denote the
different twist sectors $z$ simply by counting the number of non trivial 
twists in the various directions. Since as anticipated tunnelling among them 
is strongly suppressed, one simply
needs to choose appropriate initial conditions and employ a local update 
algorithm (standard Metropolis) within phase II to keep the twist sectors 
fixed.

A further observable which can be measured at fixed twist is
the Pisa disorder operator \cite{DiGiacomo:1997sm},
motivated by the dual superconductor scenario for the QCD vacuum 
\cite{Nambu:1974zg,Mandelstam:1975hb,'tHooft:1974qc}.
Its construction in the 
case of the $SO(3)$ at non trivial twist follows the same lines as in the 
trivial twist case \cite{Barresi:2004qa}.
The magnetically charged operator $\mu$
shifts the quantum field at a given time slice by a classical 
external field corresponding to an Abelian monopole, with the
$U(1)$ subgroup of the gauge group, which defines the magnetic charge,
selected by an Abelian projection usually fixed by diagonalizing 
an operator $X$ in the adjoint representation. As in Ref.~\cite{Barresi:2004qa} 
we will work with the random Abelian projection (RAP) introduced 
in Ref.~\cite{Carmona:2001ja}. 

The disorder parameter is defined as
\begin{eqnarray} 
\langle \mu(t)\rangle = \frac{\int (DU)_M e^{-S_M(t)}}{\int (DU) e^{-S}}\,,
\label{pisa}
\end{eqnarray}
where $S_M(t)$ denotes the Wilson action with the space-time plaquettes 
$U_{i4}(\vec{x},t)$ at a fixed time-slice $t$ modified by an insertion of 
an external monopole field 
\begin{eqnarray}
\widetilde{U}_{i4}(\vec{x},t)&=&U_i(\vec{x},t)\Phi_i(\vec{x}+\hat{i},\vec{y})
U_4(\vec{x}+\hat{i},t)\times \nonumber\\ 
&\times&U_i^{\dagger}(\vec{x},t+1)
U_4^{\dagger}(\vec{x},t))\,, 
\label{plaq_shift}
\end{eqnarray}
where
$\Phi_i(\vec{x},\vec{y})=\Omega e^{i T_a b^a_i(\vec{x}-\hat{i},\vec{y})}
\Omega^\dagger $, with
$\Omega$ the gauge transformation which diagonalizes the operator $X$. 
$T_a$ denote the generators of the Cartan subalgebra
and $\vec{b}$ the discretized transverse field generated at
the lattice spatial point $\vec{x}$ by a magnetic monopole
sitting at $\vec{y}$. 
It should be stressed that only the plaquette contribution to the 
action (\ref{ouraction}) is modified by the insertion of the monopole 
field and not the chemical potential term $\lambda$.
From the definition of $\mu$ it can be shown that
a monopole field is added at time slice $t+1$ by using
a suitable change of variable. Iterating the procedure it can
be proved that $\mu$ effectively corresponds to an operator which 
at time slice 
$t$ creates a monopole propagating forward in time
until it is annihilated by an antimonopole at $t+\Delta t$. The 
correlation function
$D(\Delta t)=\langle
\bar{\mu}(\vec{y},t+\Delta t)\mu(\vec{y},t)\rangle $
describes the creation of a monopole at $(\vec{y},t)$
and its propagation from $t$ to $t+\Delta t$. At large $\Delta t$, by cluster
property $D(\Delta t)\simeq A\exp(-M \Delta t)+\langle\mu\rangle^2 $.
$\langle\mu\rangle\neq 0$ indicates spontaneous breaking
of the $U(1)$ magnetic symmetry and hence dual superconductivity.
In the thermodynamic limit one expects $\langle\mu\rangle\neq 0$ for
$T<T_c$, while $\langle\mu\rangle = 0$ for $T>T_c$, 
if the deconfining phase transition is associated with a transition
from a dual superconductor to the normal state.
At finite temperature there is no way to put
a monopole and an antimonopole at large distance along the $t$-axis 
as it is done at $T=0$, since at $T\sim T_c$ the temporal 
extent $N_{\tau} a$ is comparable to the correlation length. Therefore, 
one measures directly $\langle \mu \rangle$ but with $C^*$-periodic 
b.c. in time direction imposed to the numerator in 
Eq. (\ref{pisa}) in order to ensure magnetic charge conservation,
$U_i(\vec{x},N_{\tau})=U_i^*(\vec{x},0)$,
where $U_i^*$ is the complex conjugate of $U_i$, in the following
indicated by a suffix 
$M$ in the observables. They effectively create a dislocation with 
magnetic charge -1 at the boundary which annihilates the positive 
magnetic charge created by the operator $\mu$. 
An analogous condition holds also for link variables defined in the 
adjoint representation, i.e.
$U_i(\vec{x},N_{\tau})=
(\mathbb{I}_3+2T^2_2)U_i(\vec{x},0)(\mathbb{I}_3+2T^2_2)$;
charge conjugation is realized in both representations
through rotations by an angle $\pi$ around the color 2-axis.
The adjoint representation makes moreover clear how $C^*$ b.c. are up
to a gauge transformation equivalent to twisted b.c. and therefore ``natural''
in our adjoint theory.
 
A technical difficulty is that, 
since $\langle\mu\rangle$ is the average of the exponential of a sum over the 
physical volume, it is affected by large fluctuations which make it 
difficult to measure in Monte Carlo simulations.
A way out is to compute the derivative
with respect to the coupling parameter $\beta$ (i.e. $\beta_A$)
$\rho=\frac{d}{d\beta}\log\langle\mu\rangle=
\langle \Pi \rangle_S - \langle \Pi_M \rangle_{S_M}$,
which yields all the relevant informations on $\mu$.
It is the difference between the Wilson plaquette action
term $\Pi$ averaged with the usual measure and the modified plaquette action
term $\Pi_M$ averaged with the modified measure
$(DU)_M e^{-S_M} / \int (DU)_M e^{-S_M}$.
The order parameter itself can in principle be reconstructed from
$\langle\mu\rangle =\exp\left(\int_0^\beta \rho(\beta')d\beta'\right)$.
$\rho$ should vanish in the thermodynamical limit for $\beta < \beta_c$
if $\langle \mu \rangle \neq 0$.
A sharp negative peak for $\rho$ diverging in the thermodynamical
limit should signal a phase 
transition associated with the restoring of the dual magnetic symmetry, while
above $T_c$ $\rho$ should show negative plateaus diverging with the volume to 
ensure $\langle \mu \rangle = 0$.

\section{Symmetry transformation}
\label{sec4}

The absence of a ``cheap'' order parameter like 
$L_F$
has been one of the major obstacles in determining the properties of
$SO(3)$ at fixed twist \cite{Datta:1999np,Barresi:2003jq,Barresi:2004qa}.
We solve here this problem by explicitly constructing the mapping
suggested in Ref.~\cite{deForcrand:2002vs} between the $SO(3)$ theory at fixed twist
and configurations classified by some positive plaquette model 
\cite{Fingberg:1995ut}.
The constraint $\sigma_c=1$ identically satisfied in phase II is key for the 
existence of such mapping: in spite of the center blindness of 
action~(\ref{ouraction}), it makes the signs of the fundamental plaquettes 
no more completely random. A further constraint is given by the value of
$z_{\mu\nu}$, where all parallel planes concurring to it in 
eq.~\ref{deftwi} must be equal due to the $\sigma_c=1$ condition. 
In the case of trivial twist, for example, this will force first of all
every 3-cube to have an even number of negative plaquettes, which will 
therefore be either parallel or will have a link in common. Furthermore,
every 2-d plane must also have an even number of negative plaquettes.
Therefore, the allowed configurations must consist of a superposition
of two possible situation: an even number
of negative stacks of plaquettes $P_{\mu\nu}$
in all parallel planes $\mu \nu$; or 
a set of negative plaquettes $P_{\mu\nu}$, $P_{\mu\rho}$ joined by a common 
link $U_{\mu}$, which must always occur in pairs in both ${\mu\nu}$ and
${\mu\rho}$ planes. In the case of non-trivial twist the situation
is similar, with the only difference that the number of negative 
plaquettes in certain planes will now be odd.

\begin{figure}[ht]
\begin{center}
\subfigure[$xyz$ at ${t}=1$]{
\includegraphics[angle=0,width=0.30\textwidth]{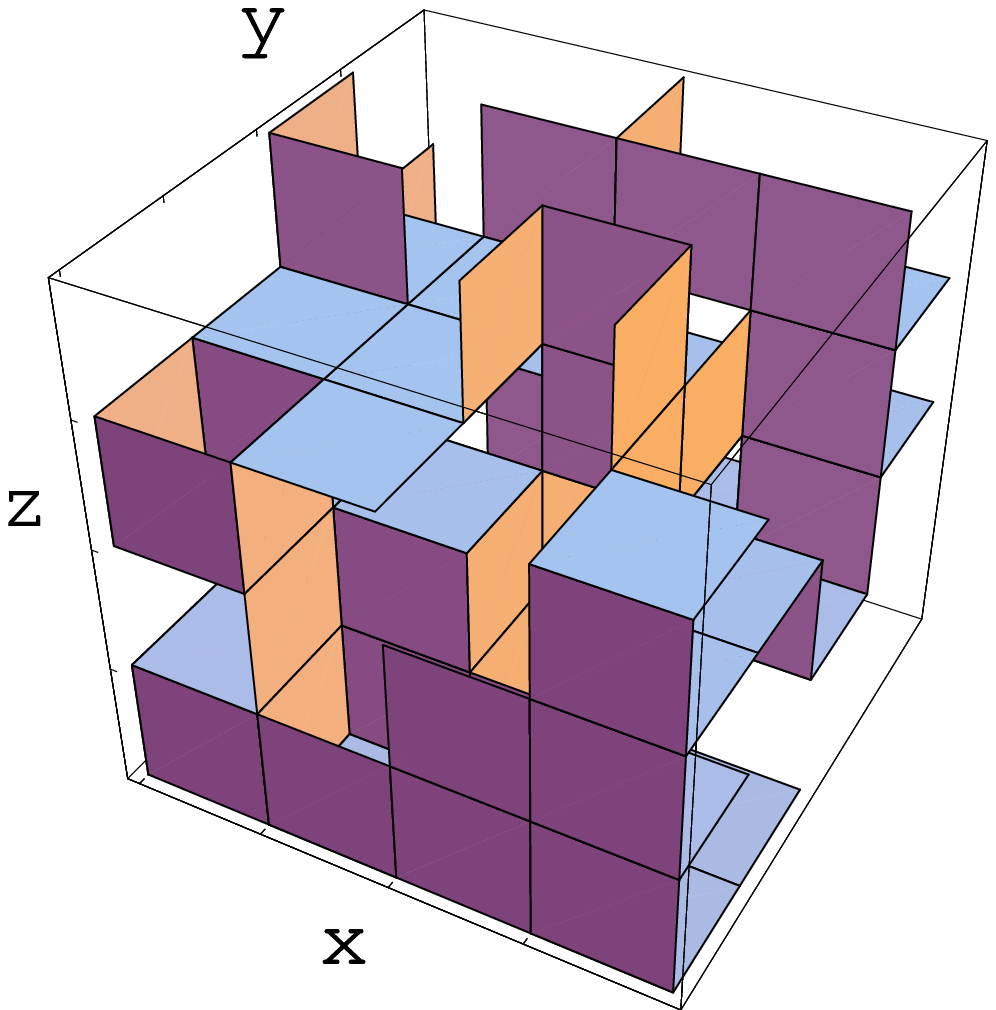}}
\subfigure[$xyt$ at ${z}=1$]{
\includegraphics[angle=0,width=0.30\textwidth]{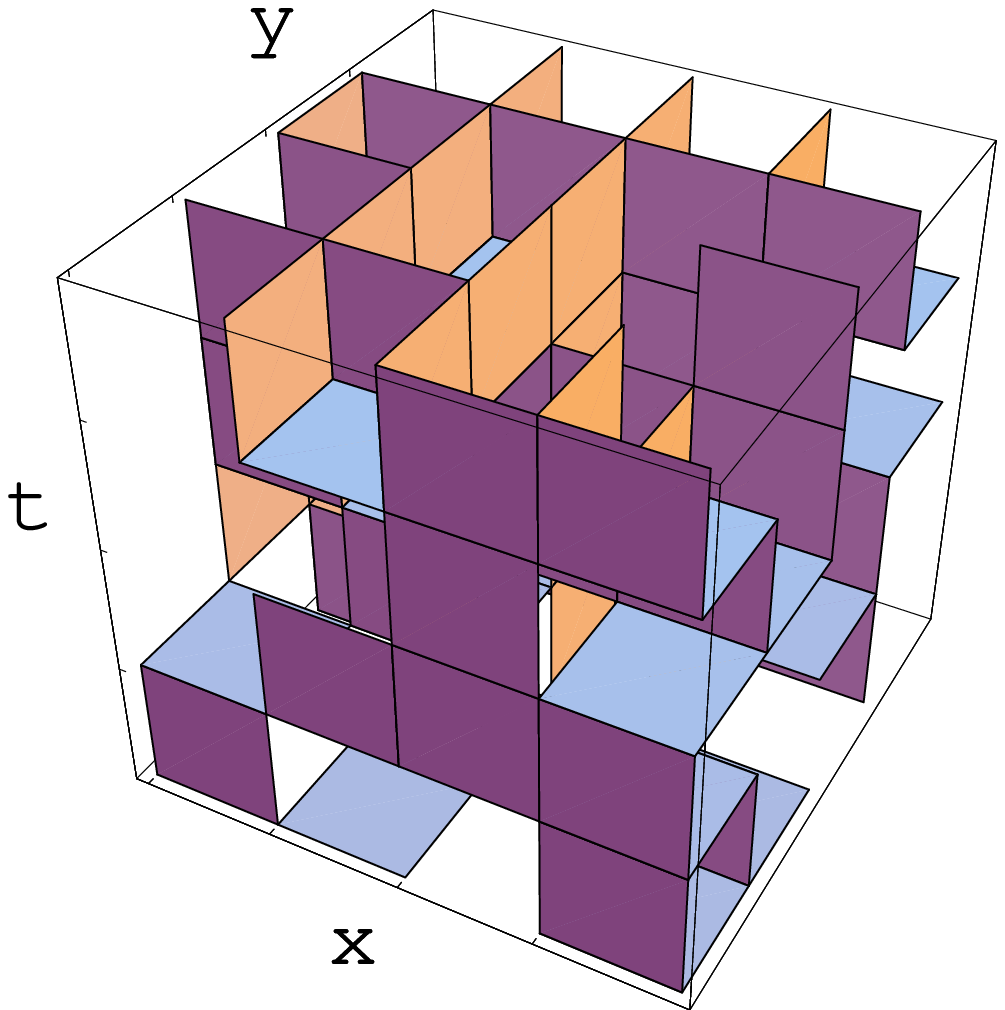}}
\subfigure[$yzt$ at ${x}=1$]{
\includegraphics[angle=0,width=0.30\textwidth]{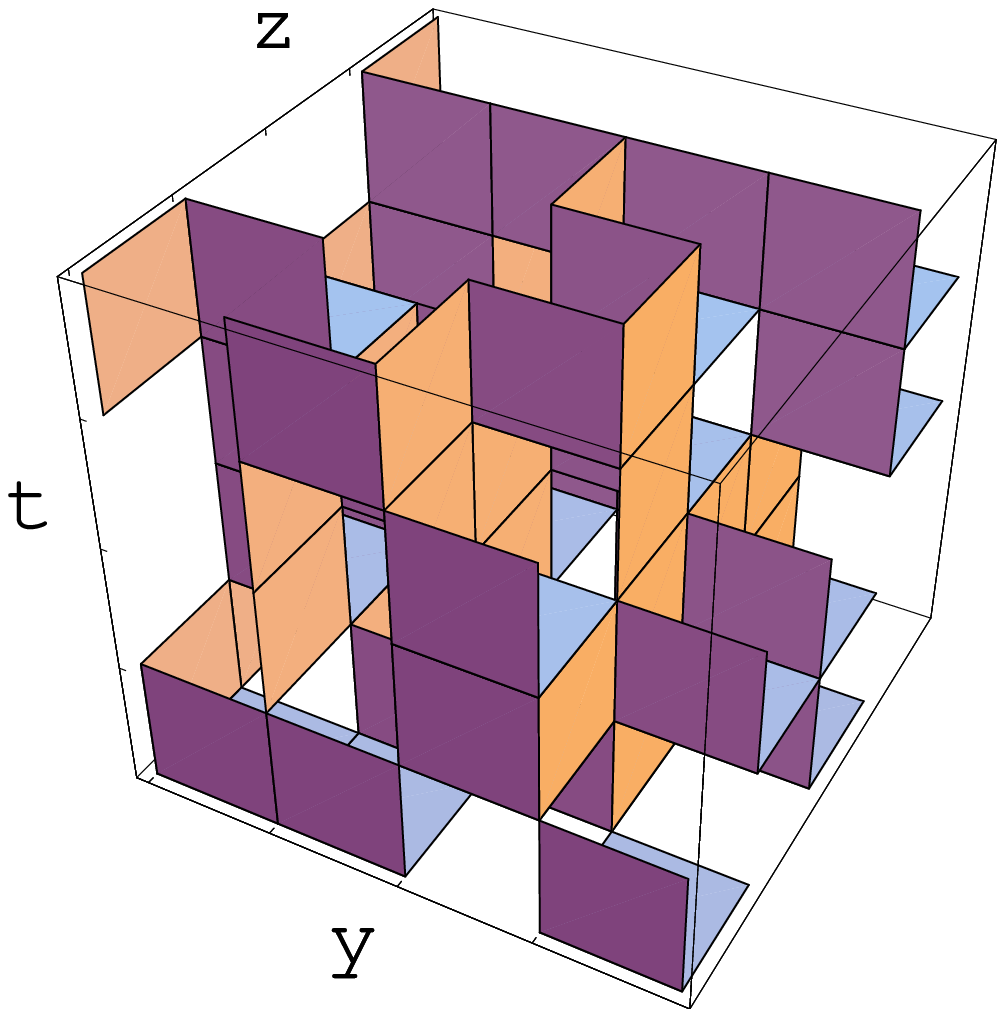}}
\subfigure[$xyz$ at ${t}=1$]{
\includegraphics[angle=0,width=0.30\textwidth]{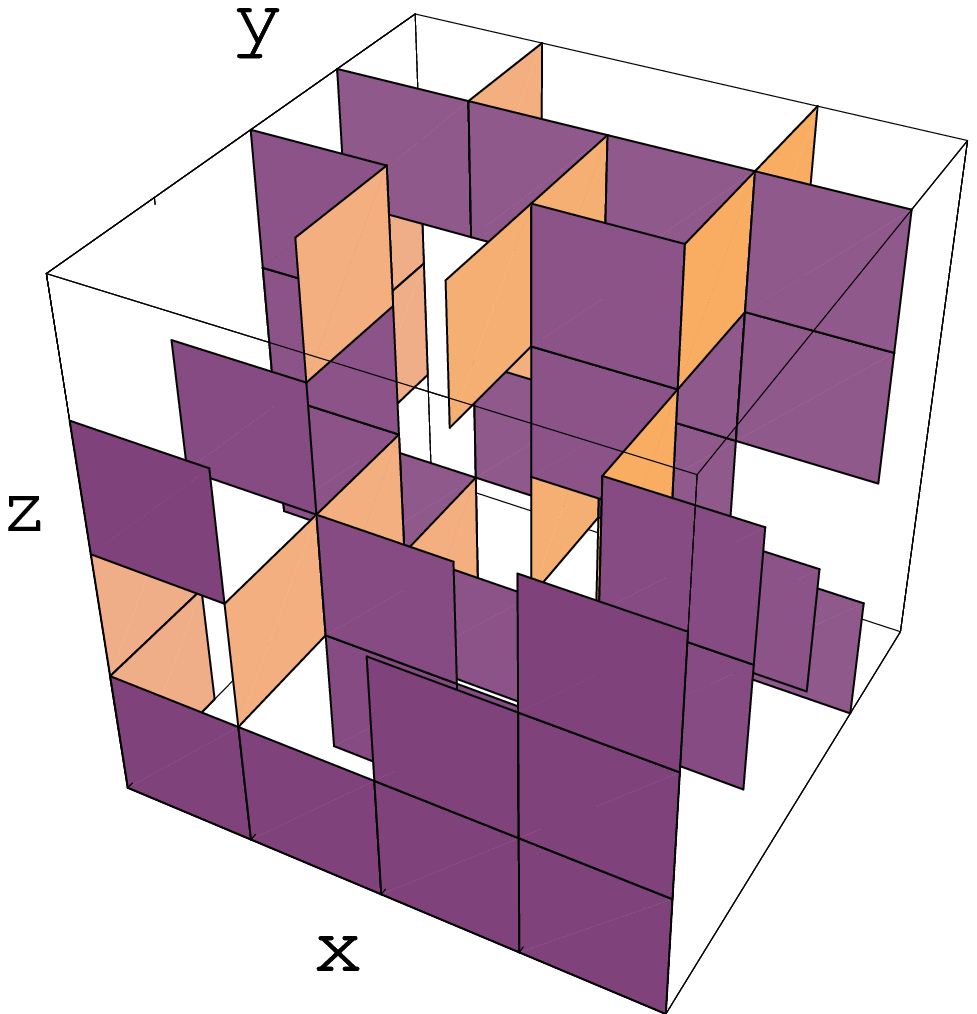}}
\subfigure[$xyt$ at ${z}=1$]{
\includegraphics[angle=0,width=0.30\textwidth]{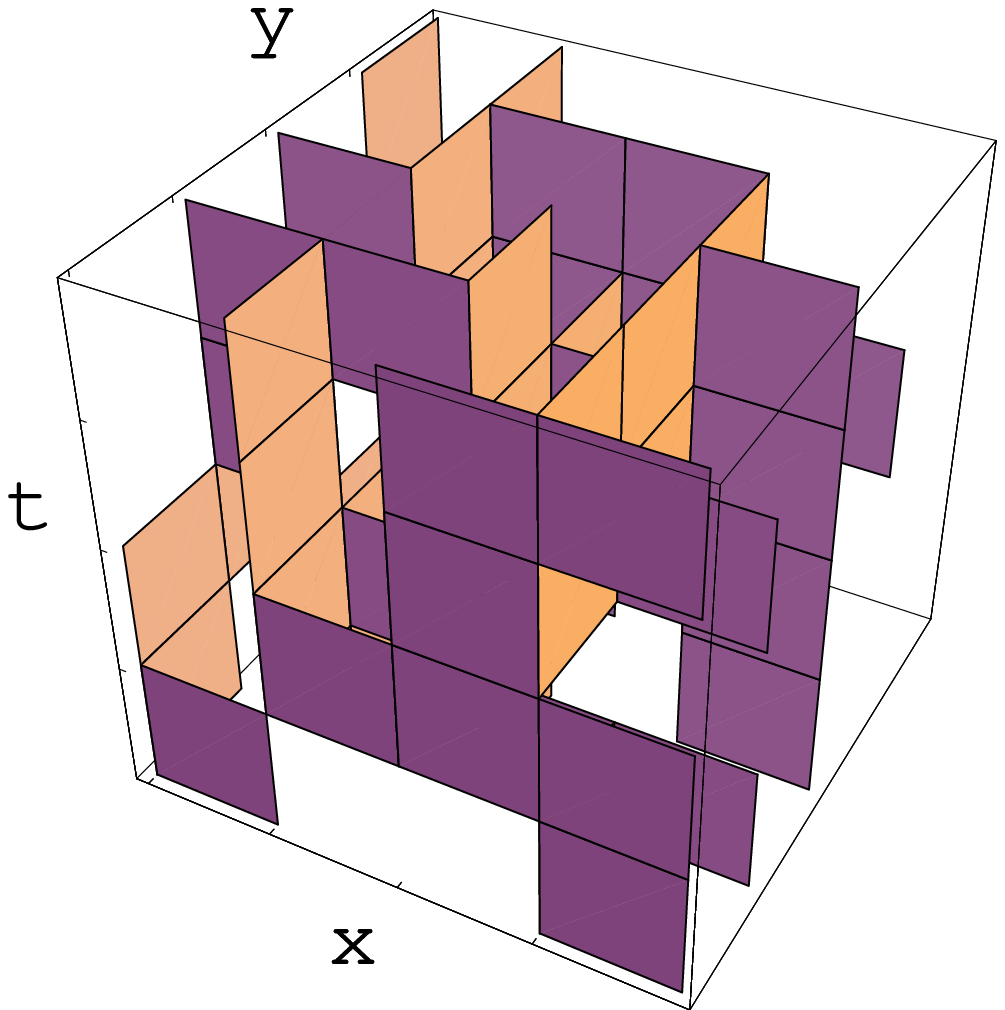}}
\subfigure[$yzt$ at ${x}=1$]{
\includegraphics[angle=0,width=0.30\textwidth]{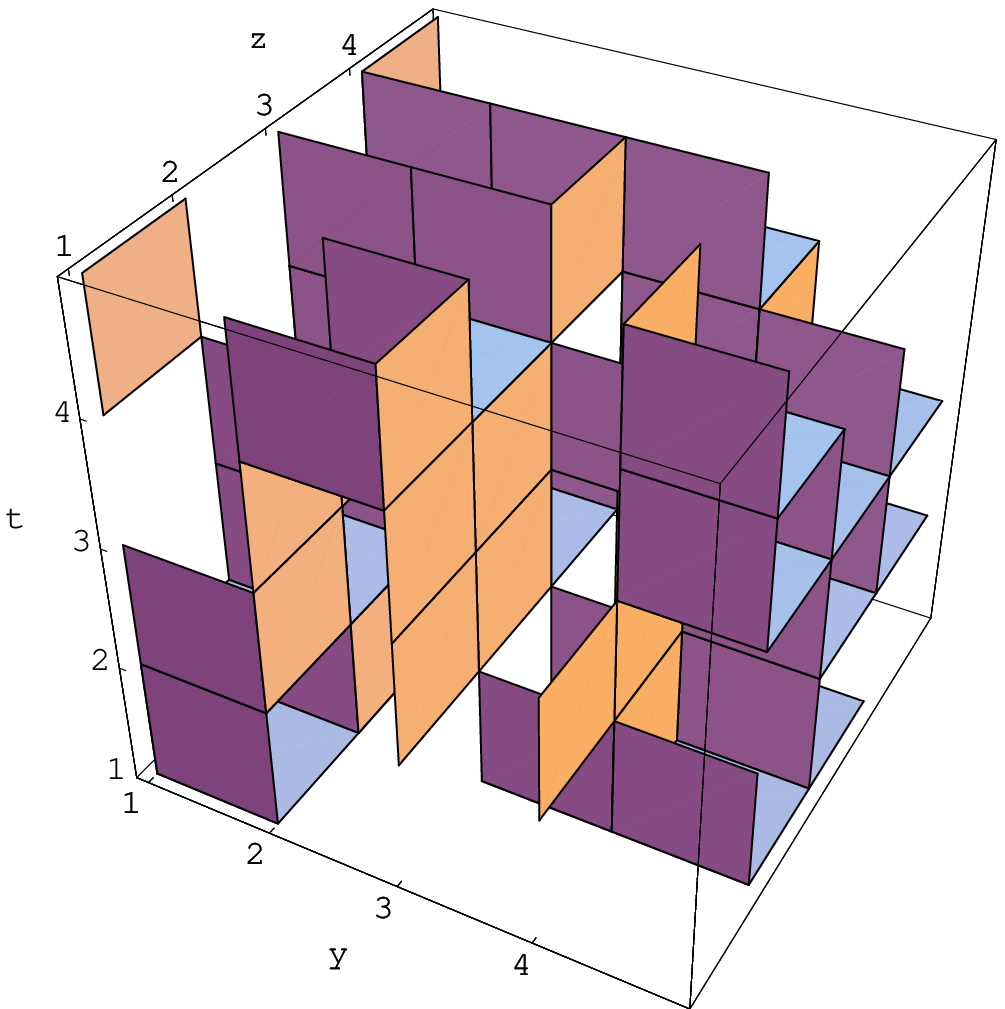}}
\subfigure[$xyz$ at ${t}=1$]{
\includegraphics[angle=0,width=0.30\textwidth]{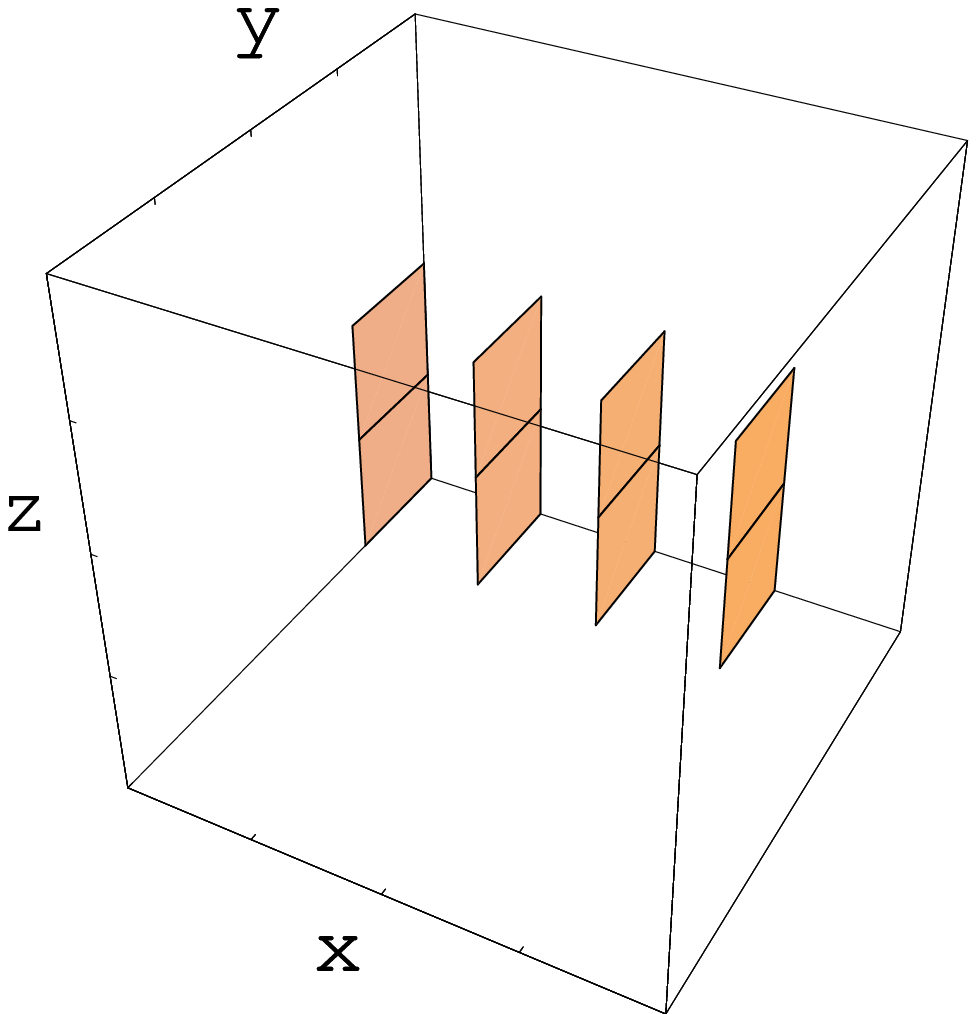}}
\subfigure[$xyt$ at ${z}=1$]{
\includegraphics[angle=0,width=0.30\textwidth]{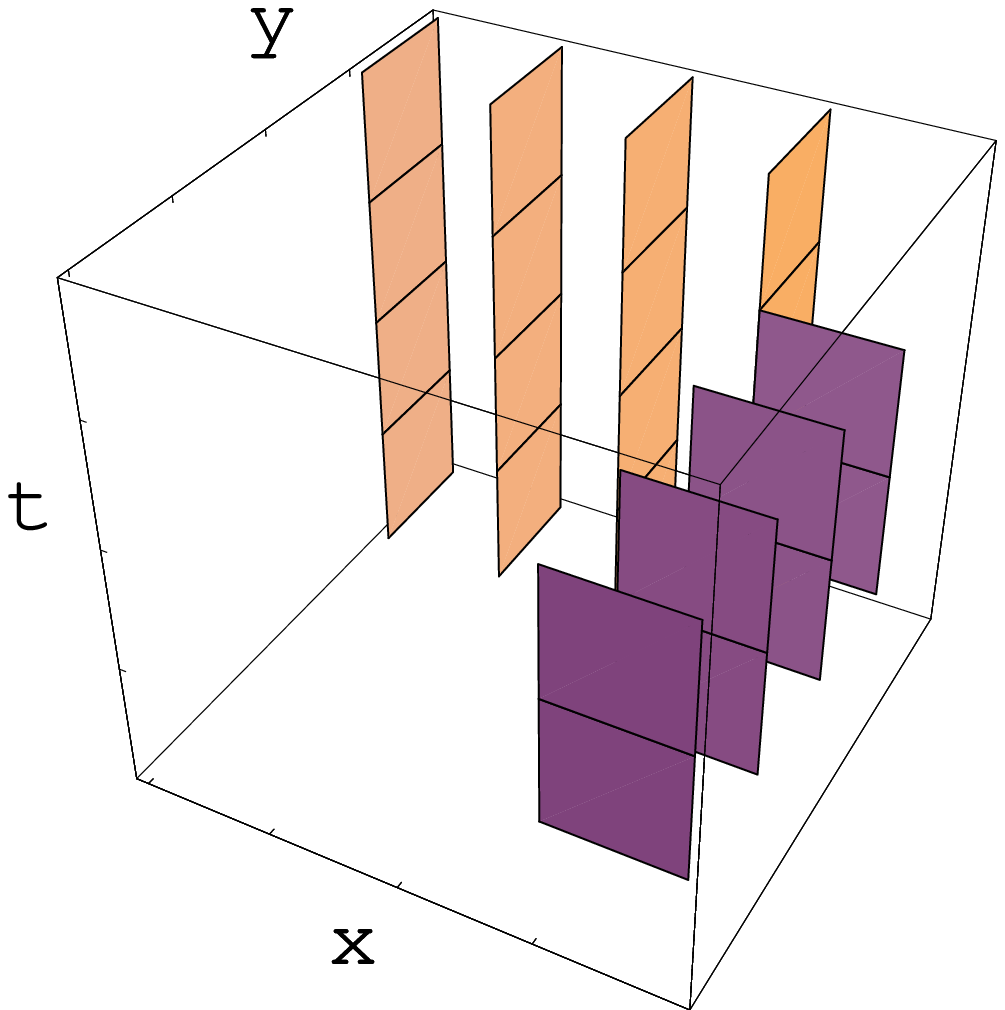}}
\subfigure[$yzt$ at ${x}=1$]{
\includegraphics[angle=0,width=0.30\textwidth]{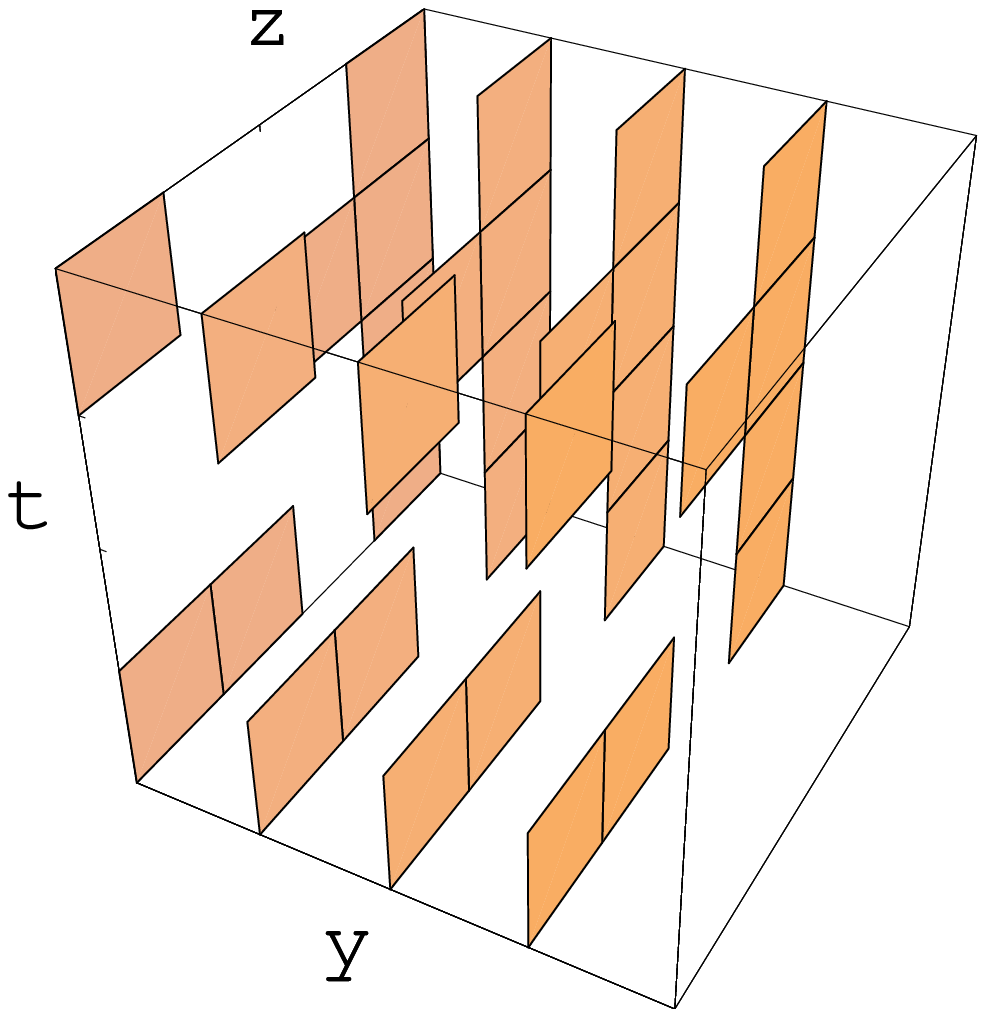}}
\end{center}
\caption{Suppression of the negative plaquettes in the trivial twist sector.}
\label{monopole}
\end{figure}

\begin{figure}[ht]
\begin{center}
\subfigure[$xyz$ at ${t}=1$]{
\includegraphics[angle=0,width=0.30\textwidth]{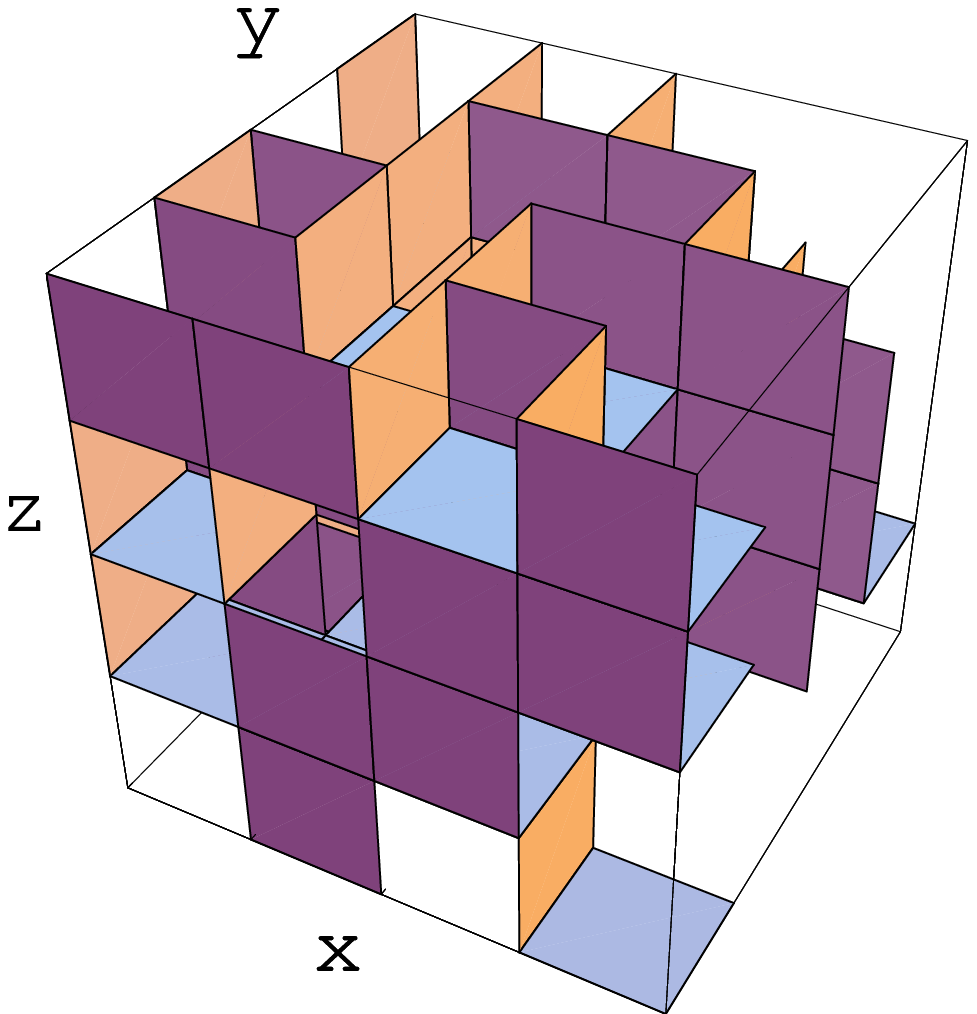}}
\subfigure[$xyt$ at ${z}=1$]{
\includegraphics[angle=0,width=0.30\textwidth]{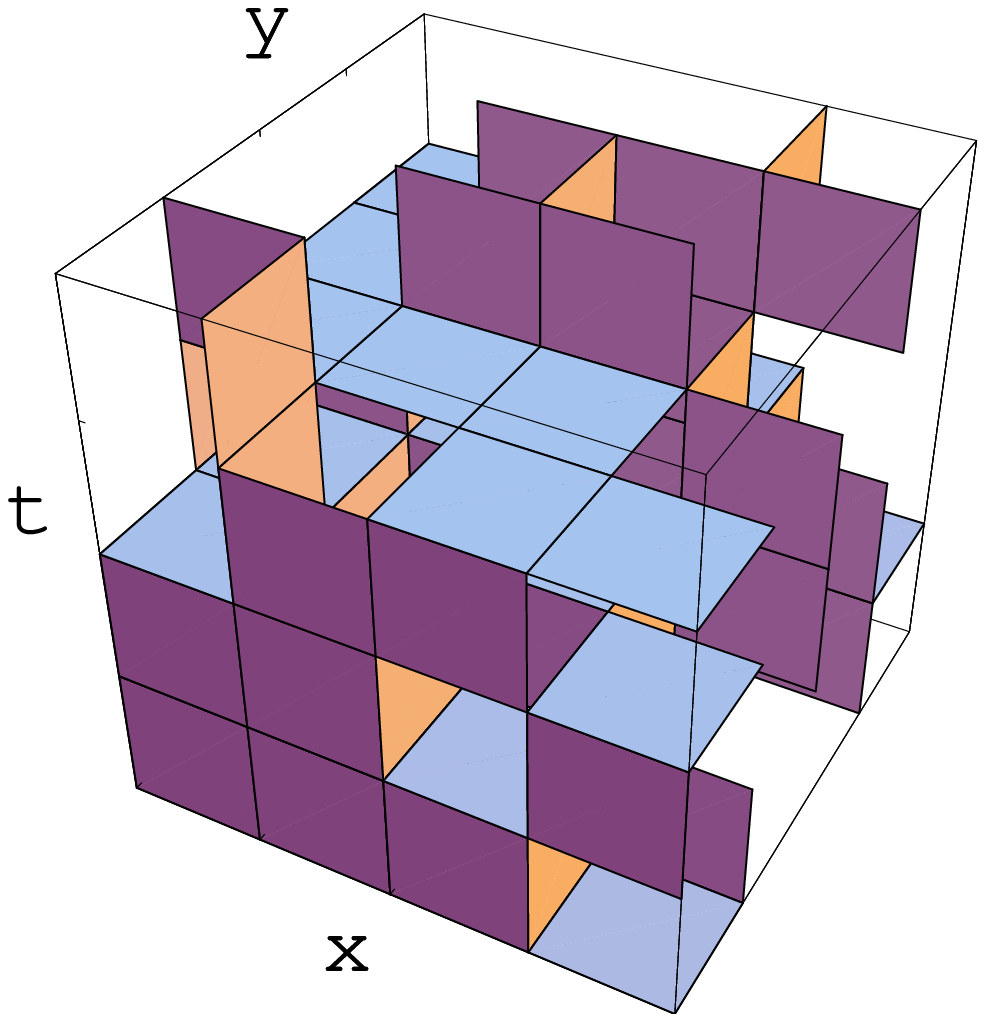}}
\subfigure[$yzt$ at ${x}=1$]{
\includegraphics[angle=0,width=0.30\textwidth]{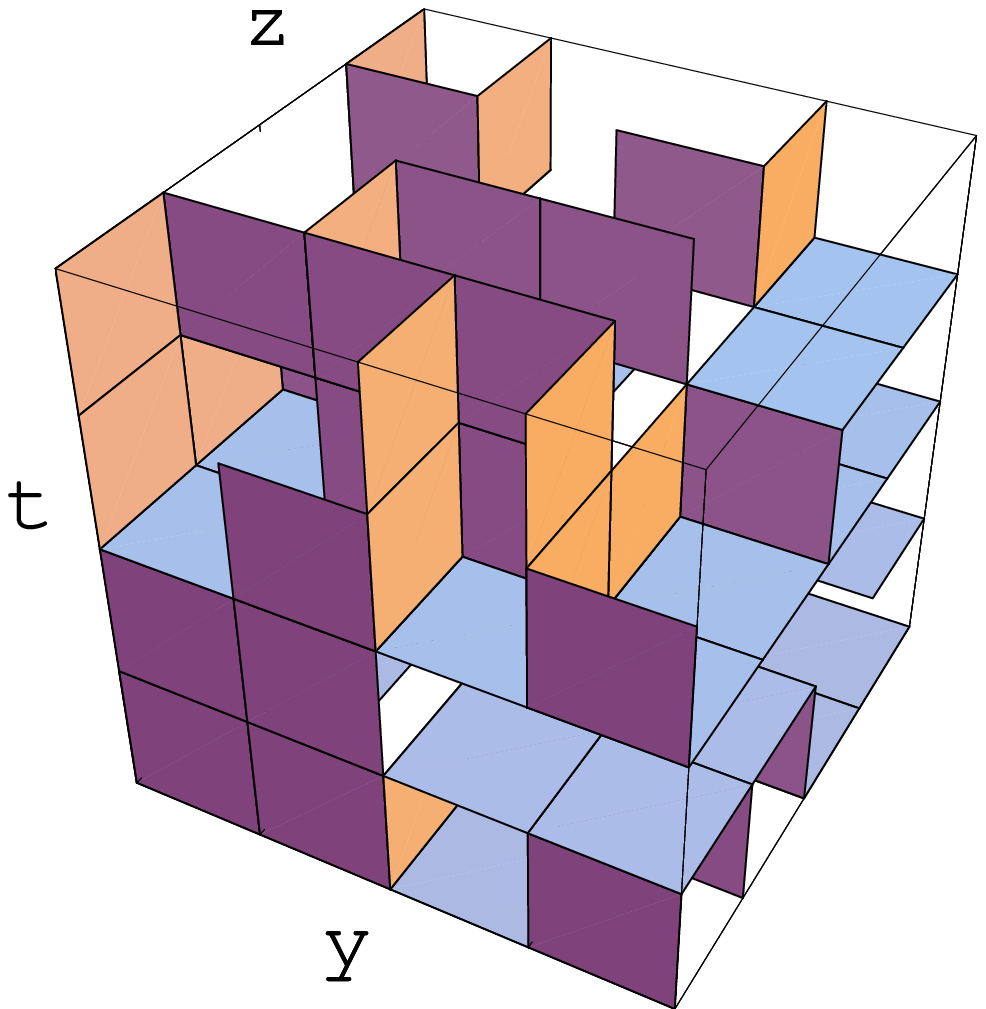}}
\subfigure[$xyz$ at ${t}=1$]{
\includegraphics[angle=0,width=0.30\textwidth]{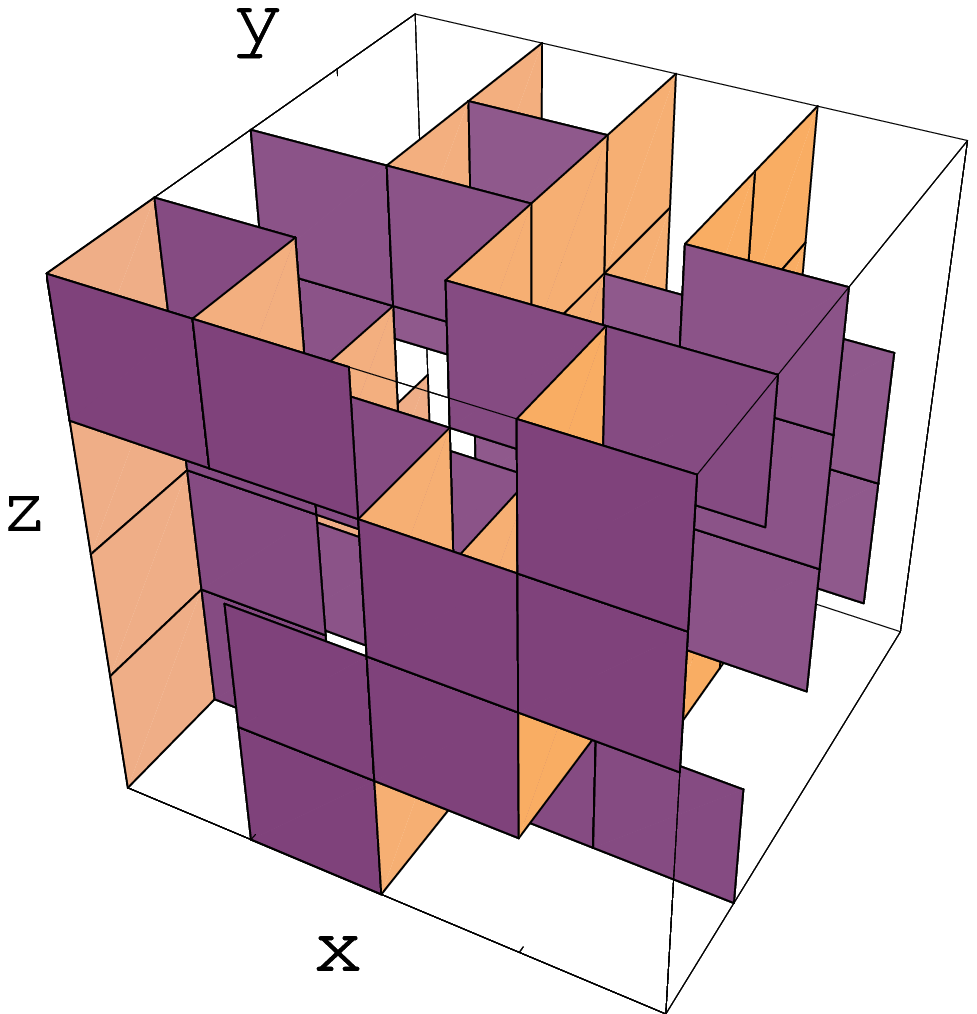}}
\subfigure[$xyt$ at ${z}=1$]{
\includegraphics[angle=0,width=0.30\textwidth]{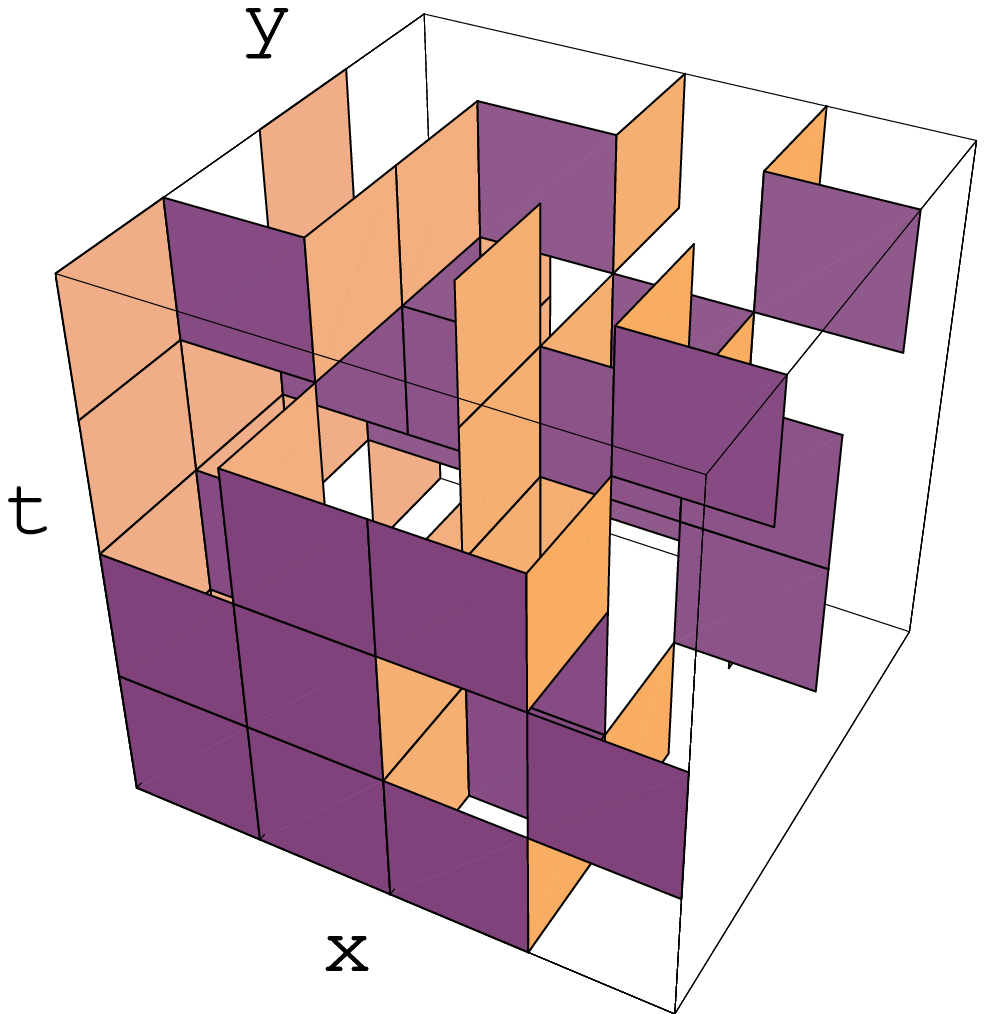}}
\subfigure[$yzt$ at ${x}=1$]{
\includegraphics[angle=0,width=0.30\textwidth]{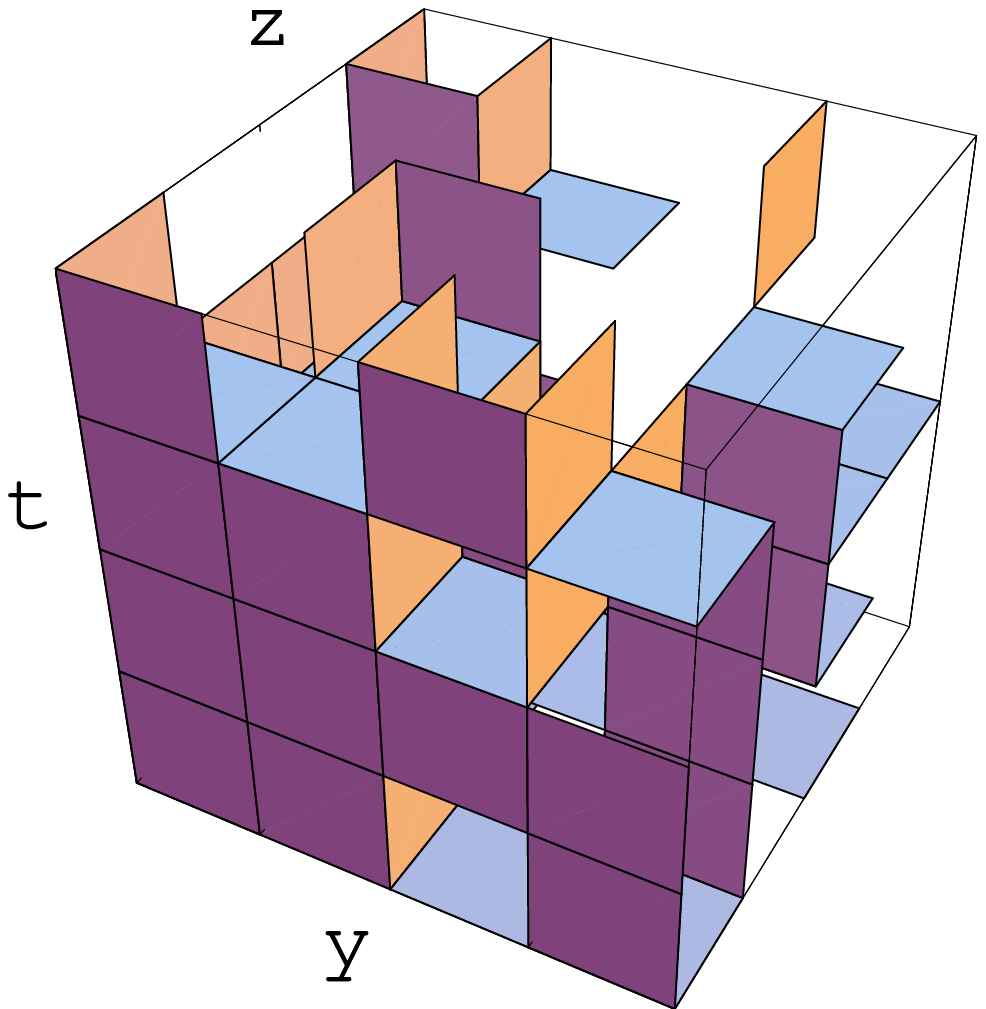}}
\subfigure[$xyz$ at ${t}=1$]{
\includegraphics[angle=0,width=0.30\textwidth]{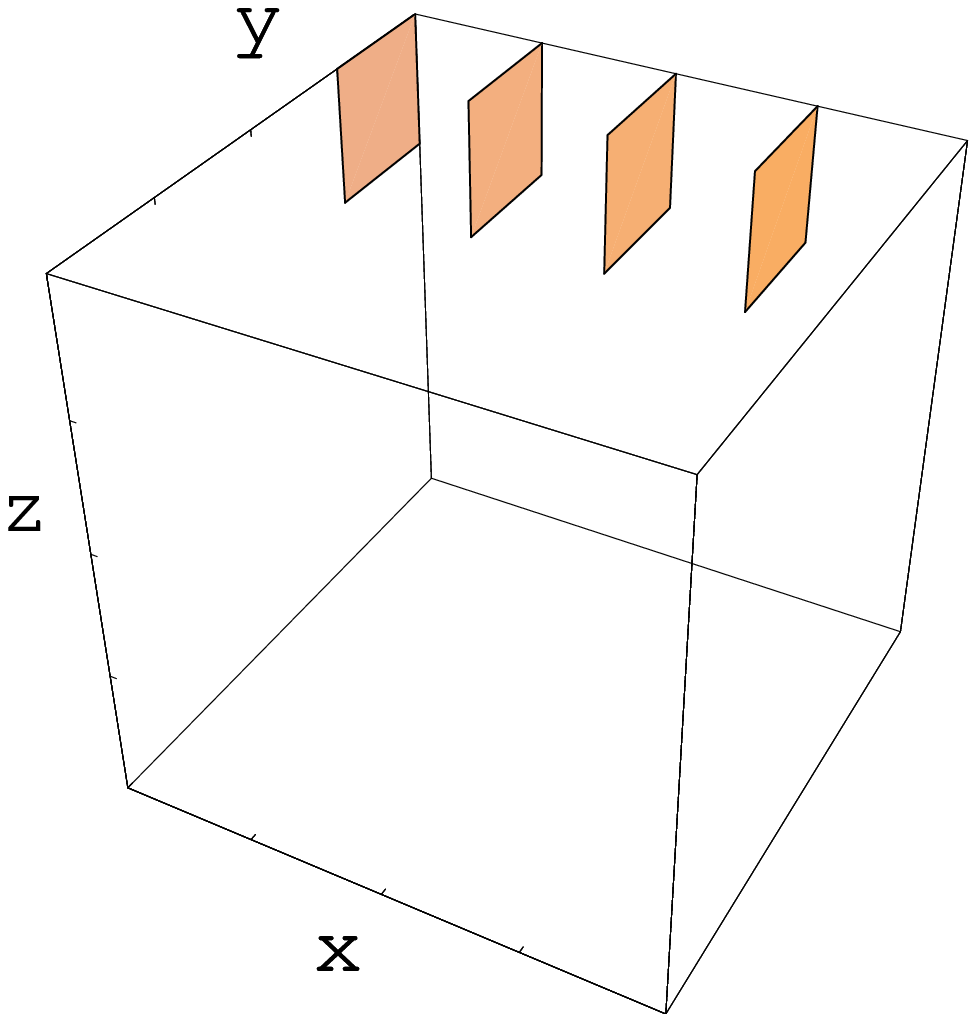}}
\subfigure[$xyt$ at ${z}=1$]{
\includegraphics[angle=0,width=0.30\textwidth]{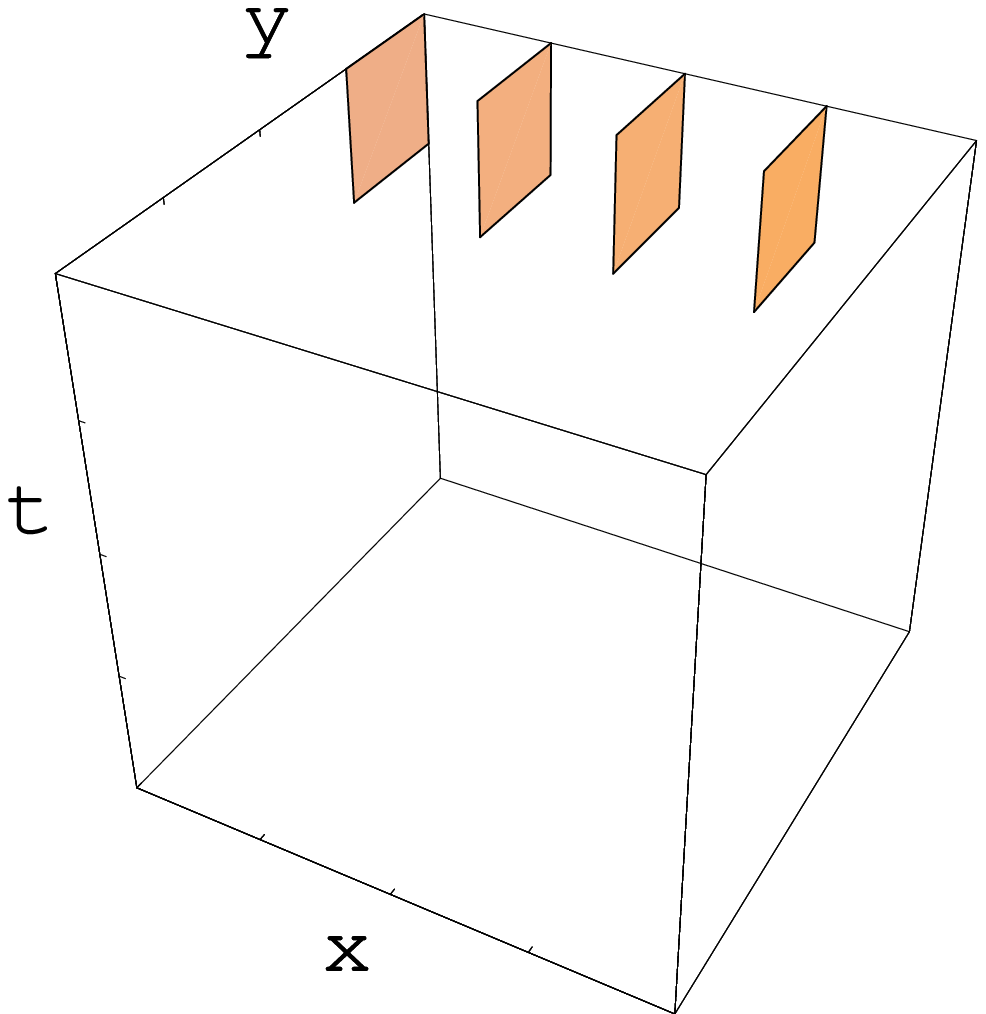}}
\subfigure[$yzt$ at ${x}=1$]{
\includegraphics[angle=0,width=0.30\textwidth]{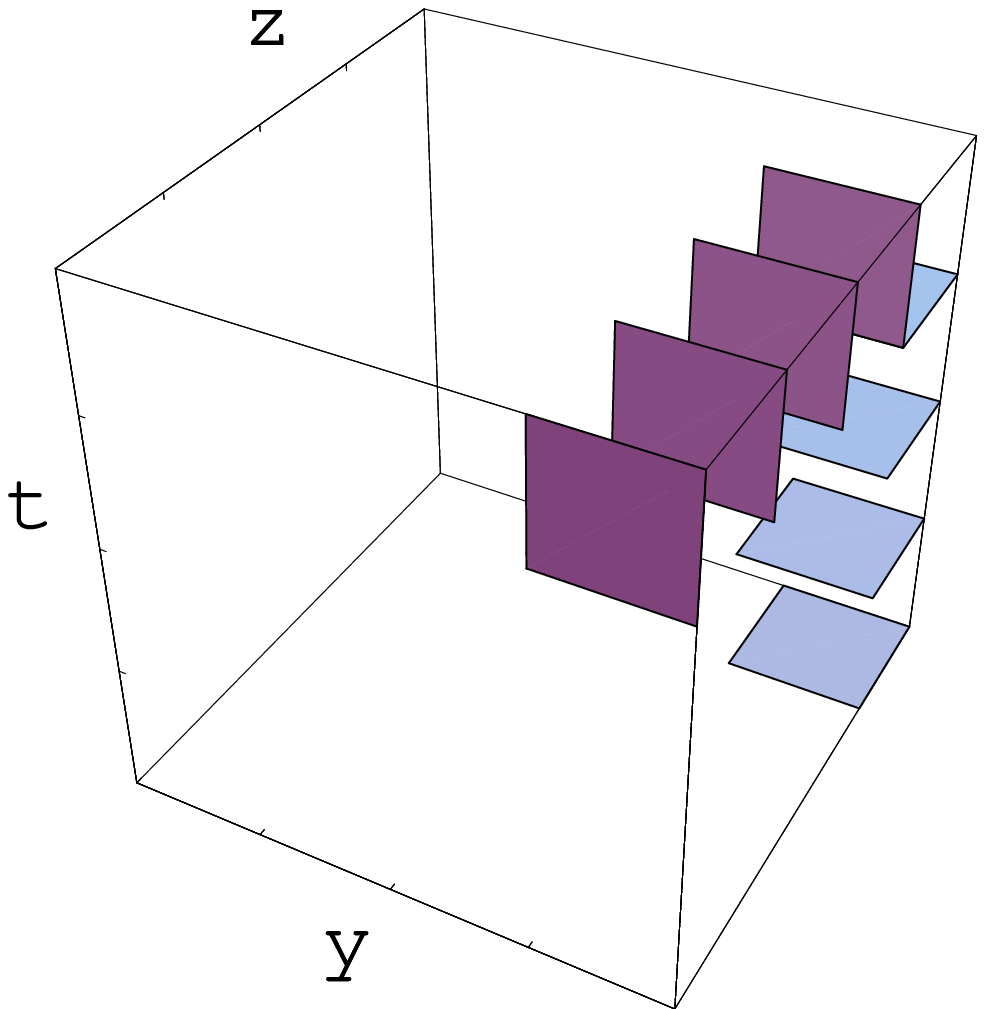}}
\end{center}
\caption{Suppression of the negative plaquettes in the non-trivial twist sector.}
\label{montwist}
\end{figure}

As an illustration, take a configuration arising from a simulation at $N_s = N_\tau=4$
(Fig.~\ref{monopole} (a-c), Fig.~\ref{montwist} (a-c)). 
By flipping the sign of a generic link six plaquettes will 
be affected while obviously $\sigma_c$ and $z_{\mu\nu}$
will remain unchanged. One can therefore start to sweep the whole
lattice changing first the sign of some links so to make all plaquettes 
in the $xy$ plane positive (Fig.~\ref{monopole} (d-f), Fig.~\ref{montwist} 
(d-f)). One can now proceed with a second sweep which makes all
plaquettes in the $xz$ plane positive by leaving the $xy$ untouched, i.e.
by flipping only $z$ and $t$ links. Fig.~\ref{monopole} (g-i), 
Fig.~\ref{montwist} (g-i) now illustrate clearly the situation: only pairs
of plaquettes which can be made positive by a $t$ link flip are present.
For non trivial twist the procedure will be the same, with the only difference 
that of course at the end a stack of plaquettes ensuring $z_{\mu\nu}=-1$,
i.e. twisted b.c., must remain.
Such procedure provides us with a kinematic identification of adjoint
configurations with a positive plaquette model (with periodic or twisted 
b.c.); it is not a dynamical mapping to a positive plaquette action of the 
type given in Ref.~\cite{Fingberg:1995ut}, since we still generate our fields 
with an adjoint center-blind weight. In some sense it amounts to
a ``gauge fixing'' which removes the local $\mathbb{Z}_2$ freedom 
intrinsic to the adjoint weight; another choice, e.g. all negative 
plaquettes, could have been equally made. 
Since no dynamical identification is possible,
the positive plaquette fields we obtain will not be identical to the one
generated through the action given in Ref.~\cite{Fingberg:1995ut}.
They might however exhibit similar scaling properties.
For each configuration generated in the MC with weight given by 
eq.~\ref{ouraction} we have applied the above algorithm to obtain a 
configuration where fundamental observables do not vanish identically when 
averaged over the volume, although they should be interpreted as 
gauge dependent observables in a gauge fixed theory. 

We can now measure $L_F$ and its susceptibility
in all fixed twist sectors of the adjoint theory
and use it as an order parameter to determine the critical exponents.
Although in Ref.~\cite{deForcrand:2001nd} an alternative
definition of $L_F$ modified via a twist eater at the boundary
has been used for the fundamental
theory with fixed twisted b.c., we have chosen to stick to the
standard definition for a number of reasons.
From a purely formal point of view, for the full adjoint theory reflection
positivity can only be invoked for adjoint Polyakov loop 
correlators, ensuring their positivity irrespective of
the twist sector; in other words such quantities are invariant also
under large, twist changing gauge transformations and should be considered the
``fundamental'' ones. Since we wish to maintain the property 
$\mathrm{Tr}_{A} = 2 \mathrm{Tr}_{F}^2 - 1$ also for the local
Polyakov loop $L(\vec{x}) = \prod_{t=0}^{N_{\tau}} U_4(\vec{x},t)$, so to
keep the volume average of $L_A$ proportional to the second moment of the 
spatial distribution of $L_F$, to ensure the correct behaviour under gauge 
transformations we are forced to keep the standard definition of $L_F$.
From a practical point of view, our algorithm does not pick a particular
representative of the homotopy class ensuring the fixed b.c. corresponding 
to the twist sector chosen. One would need to further fix the single links 
throughout the lattice to have b.c. representable through some specific twist
eater at the boundary. The definition in Ref.~\cite{deForcrand:2001nd} is
therefore not easily applied to configurations derived from an adjoint
weight at fixed twist sector. We think however that there is
no real problem underlying such ambiguity, since measuring $L_F$ 
at non trivial twist is anyway an unphysical procedure.
No fermion fields, not even in the limit of infinite mass, can be
present with twisted b.c., so that no real physical interpretation
can be given to $L_F$ and its correlators at non trivial twist.

\section{Fixed twist vs. ergodic simulations}
\label{sec3}

For the adjoint theory in phase II, being the twist sectors 
$z$ well defined,
the ergodic expectation value of any observable $O$ can be always re-expressed 
through:
\begin{equation}
\langle O \rangle_{erg} = \frac{\sum_{i=0}^{3} \langle O\rangle|_{z={i}} 
Z_{SO(3)}|_{z={i}}}{\sum_{i=0}^{3} Z_{SO(3)}|_{z={i}}}\,,
\label{erg}
\end{equation}
where $\langle O\rangle|_{z={i}}$ is the expectation value of the
observable restricted to the fixed twist sector $z=i$. 

In general in absence of an ergodic algorithm the relative weights 
of the partition functions $Z_{SO(3)}|_{z={i}}$ remain unknown, although
there is of course at least one case in which Eq.~(\ref{erg}) is of use in 
a fixed twist analysis, namely when 
$\langle O\rangle|_{z={i}}\simeq \langle O\rangle$ $\forall i$, i.e. the
observable is independent of the twist sector;
in such case it will obviously be 
$\langle O \rangle_{erg}\simeq\langle O\rangle$. 

The ratio of partition functions could in principle be estimated
from the behaviour of the vortex free energy 
\cite{'tHooft:1979uj,Tomboulis:1980vt,Kovacs:2000sy,deForcrand:2000fi,deForcrand:2001nd,deForcrand:2005pb}, 
leading to 
\begin{equation}
\frac{Z_{SO(3)}|_{z={i}}}{Z_{SO(3)}|_{z={j}}} \simeq c_{ij}
\label{low}
\end{equation}
in the confined phase, while in the (deep) deconfined phase $T >> T_c$
\begin{equation}
\frac{Z_{SO(3)}|_{z={i}}}{Z_{SO(3)}|_{z={j}}} \simeq c_{ij} 
e^{-\frac{\tilde{\sigma}N_s^2}{T}(i-j)}\,,
\label{high}
\end{equation}
$N_s$ being the spatial length of the lattice, $\tilde{\sigma}$ the dual 
string tension and the coefficients $c_{ij}$ taking the values
\begin{eqnarray} 
c_{ji} &=& c_{ij}^{-1}; \; c_{ii}=1; \; c_{10}=c_{20}=3; \; \nonumber \\ 
c_{21} &=& c_{30}=1; \; c_{31}=c_{32}=\frac{1}{3}\,.
\end{eqnarray}  

This is quite straightforward to see: being the action cost 
to create $(i-j)$ maximal 't~Hooft loops strictly zero in $SO(3)$, 
the free energy to tunnel twist sector from $i$ to $j$ will simply be given by 
the entropy change 
\begin{equation}
\Delta F_{ij} = - T \Delta S_{ij} = 
- T \log{\frac{Z_{SO(3)}|_{z={i}}}{Z_{SO(3)}|_{z={j}}}}\,. 
\end{equation}
This on the other hand will be dictated in the deconfined (confined) phase 
by an area (perimeter) law through the dual string tension, i.e. 
$\Delta F_{ij} = 0$ for $T<T_c$ and 
$\Delta F_{ij} = T \tilde{\sigma}N_s^2 (i-j)$ for $T>>T_c$. 
The factors $c_{ij}$ are due to the 
counting of states on a 3-torus topology, as indeed the existence of 
twists states with $z>1$ in the first place. In other words, 
on $T^3 \times S^1$ there is 
one $z=0$ twist state, three independent $z=1$ and $z=2$ states and one $z=3$
state, while on $S^3 \times S^1$ there is only one $z=0$ and one $z=1$ 
twist state. 

In the thermodynamic limit $N_s \to \infty$ all twist states should therefore 
be equivalent in the confined phase with
\begin{equation} 
\langle O \rangle_{erg} = \frac{\sum_{i=0}^{3} \langle O\rangle|_{z={i}} c_{i0}}{8} 
\label{mintc}
\end{equation} 
while all the non-trivial ($z > 0$)
states will be exponentially suppressed above $T_c$ with 
\begin{equation} 
\langle O \rangle_{erg} = (1+e^{-\frac{\tilde{\sigma}N_s^2}{T}})^{-3} 
\sum_{i=0}^{3} \langle O\rangle|_{z={i}} c_{i0} 
e^{-\frac{\tilde{\sigma}N_s^2}{T}i},
\label{majtc}
\end{equation} 
so that if the $\langle O\rangle|_{z={i}}$ are bounded (or diverge less than 
exponentially)
for $N_s \to \infty$ then obviously 
$\langle O \rangle_{erg} \simeq \langle O\rangle|_{z={0}}$. This will
hold of course if $\langle O\rangle|_{z={i}}\simeq 0$ $\forall i \neq 0$.

There are however some difficulties with this standard picture. The vanishing 
of 
the dual string tension is just a sufficient and not a necessary condition
for the existence of a confined phase, which on the other hand
only assures that the deconfined 
behaviour in the regime $T>>T_c$ will necessarily obey eq.~(\ref{majtc}). 
No result was actually available in the literature for 
the confined phase of the ergodic $SO(3)$ until the studies in 
\cite{Burgio:2005xe,Burgio:2006dc,Burgio:2006xj}
appeared, which however point to a different behaviour
of the center blind adjoint discretization from the fundamental one. We
will therefore avoid to use eq.~(\ref{mintc}) in the interpretation of
the results.
This will however turn out not to be a major problem, most observable
of interest resulting independent of the twist sector below $T_c$.

\section{Results}

\subsection{The Pisa disorder operator}
\label{sec:pisa}

\begin{figure}[thb]
\begin{center}
\subfigure[]{
\includegraphics[angle=0,width=0.45\textwidth]{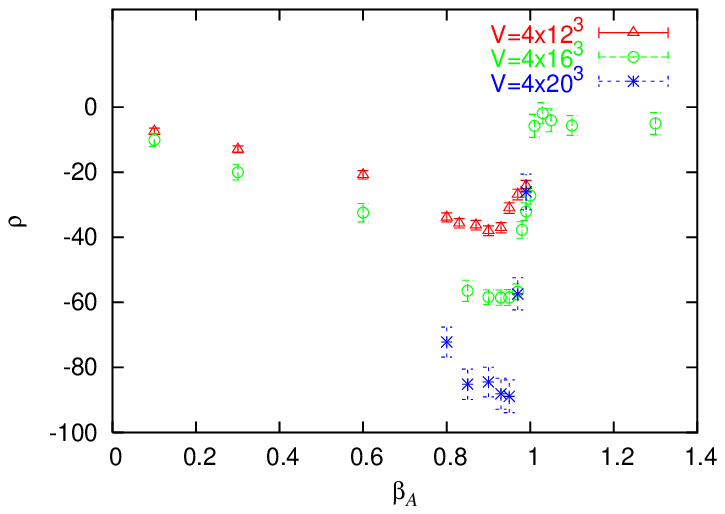}}
\subfigure[]{
\includegraphics[angle=0,width=0.45\textwidth]{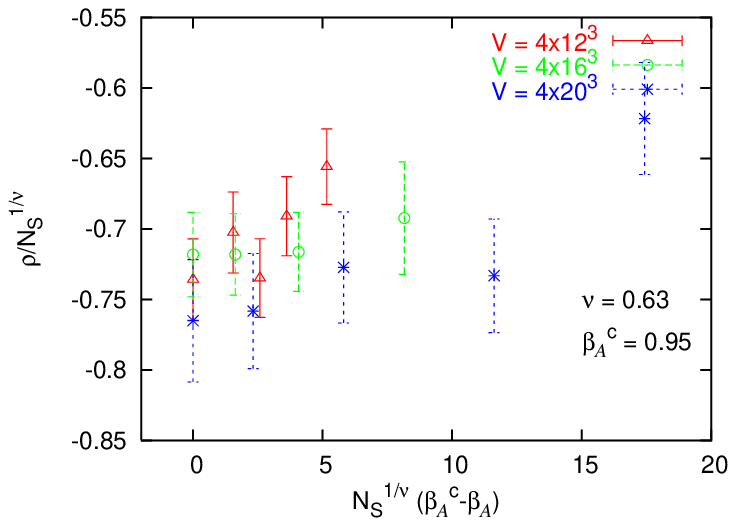}}
\end{center}
\caption{$\rho$ computed in the non trivial twist sector
at finite temperature ($N_{\tau}=4$) for different values of the
spatial volume and (a).
Finite-size scaling analysis for $\rho$ (b).} 
\vspace{1cm}
\label{nt4}
\end{figure}

The analysis of $\rho$ in the trivial twist sector has already been carried 
out in Ref.~\cite{Barresi:2004qa} for different $N_{\tau}$. In Fig.~\ref{nt4}(a)
we show the behaviour of $\rho$ for chemical potential $\lambda=1.0$ at 
$N_{\tau}=4$ in the non trivial twist sector.
The similarities with the trivial twist case end with the peak which should 
signal the transition ($\beta_A^c$ = 0.95). 
Above $\beta_A^c$ $\rho$ vanishes, in contrast to the strong diverging plateaus seen
at trivial twist \cite{Barresi:2004qa}. The transition at non trivial twist
therefore cannot correspond,
strictly speaking, to a deconfined phase. We shall try to better understand
this through the analysis of other observables. 
A consistency FSS analysis taking the value $\beta_A^c = 0.95$ and the critical 
exponents of the 3-d Ising model is shown in Fig.~\ref{nt4}(b). We will comment
on its quality in Sec.~\ref{sec:pot}.
One thing we would like to stress here 
is that such vanishing of $\rho$ poses no problem in the ergodic theory. Given 
the behaviour at trivial twist,
eq.~\ref{majtc} ensures us
that the Pisa disorder parameter will indicate deconfinement at high $\beta$,
provided that there exists a diverging peak at some $\beta_A^c$. Since the
peaks in the trivial and non trivial twist sector occur at slightly different
$\beta_A^c$ ($\beta_A^c = 0.98$ for $z=0$, cfr. \cite{Barresi:2004qa} and the
following section), this latter question can only be answered 
by a full ergodic simulation \cite{Burgio:2006xj}. The situation at low $\beta$ is 
slightly more 
complicated. Although in all twist sectors $\rho$ assumes a constant
bounded small negative value $\rho \simeq - k$ (cfr. \cite{Barresi:2004qa}), 
therefore 
indicating $\langle \mu \rangle\neq 0$ also for the full ergodic theory, i.e.
condensation of monopoles and confinement in the low $\beta$ region, the fact 
that it does not seem to strictly vanish in the thermodynamic limit might pose 
a conceptual problem: for every fixed $N_{\tau}$ $\langle \mu \rangle$ can
be redefined post-hoc to assume a constant value through 
$exp(k \beta_A^c(N_{\tau}))$, but this rescaling factor will
necessarily diverge for $SU(N)$ like $N_{\tau}^{\alpha k}$, up to logarithmic 
corrections depending on the 
higher order coefficients of the $\beta$-function, with 
$\alpha= 2 \beta_0 (N^2-1)/N$, $\beta_0$ being its first coefficient. 
This might
on the other hand be a conceptual obstacle in defining 
$\langle \mu \rangle$ in the 
continuum limit $N_{\tau} \to \infty$, although $\alpha$ is just 
$\simeq 1.4$ for $SU(2)$.

\subsection{$L_F$ and the interquark potential}
\label{sec:pot}

\begin{figure}[htb]
\begin{center}
\subfigure[]{
\includegraphics[angle=0,width=0.45\textwidth]{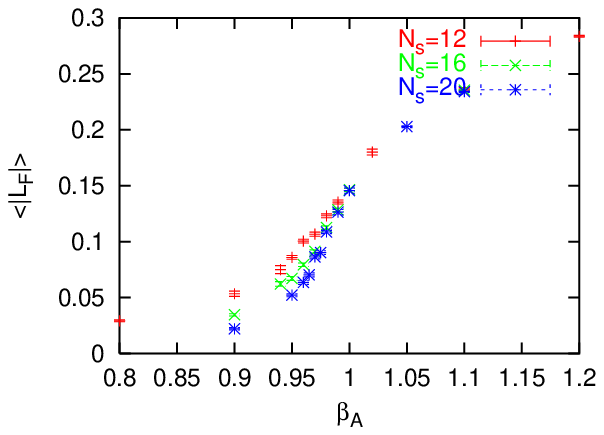}}
\subfigure[]{
\includegraphics[angle=0,width=0.45\textwidth]{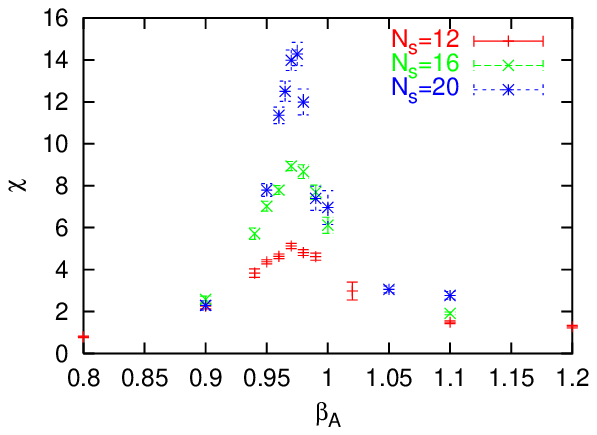}}
\end{center}
\caption{$L_F$ (a), its susceptibility $\chi$ (b).}
\label{poltr4}
\end{figure}

\begin{figure}[htb]
\begin{center}
\subfigure[]{
\includegraphics[angle=0,width=0.45\textwidth]{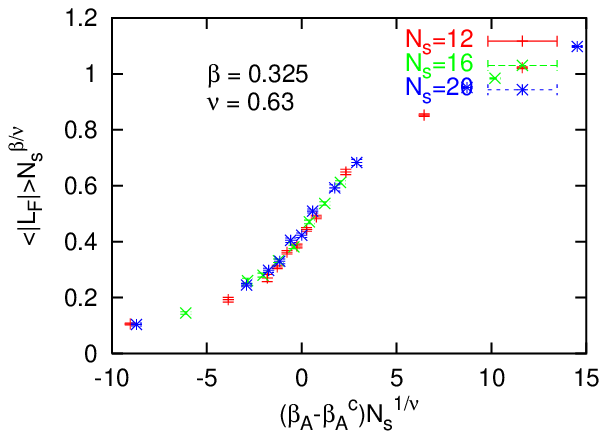}}
\subfigure[]{
\includegraphics[angle=0,width=0.45\textwidth]{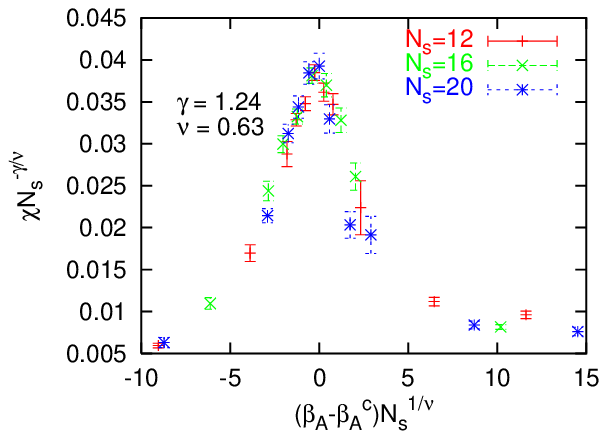}}
\end{center}
\caption{Quality of scaling for 
$L_F$ (a) and for $\chi$ (b).}
\label{poltr44}
\end{figure}

\begin{figure}[htb]
\begin{center}
\subfigure[]{
\includegraphics[angle=0,width=0.45\textwidth]{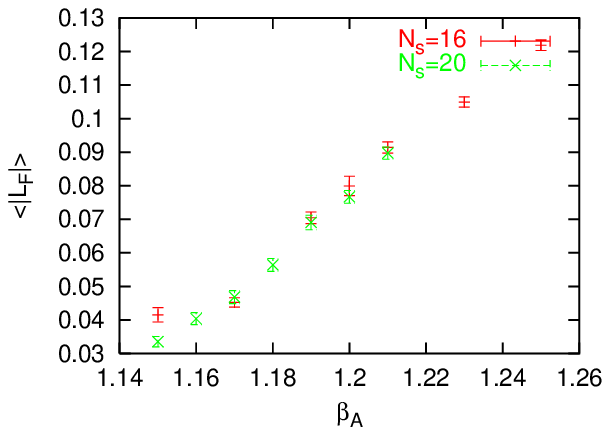}}
\subfigure[]{
\includegraphics[angle=0,width=0.45\textwidth]{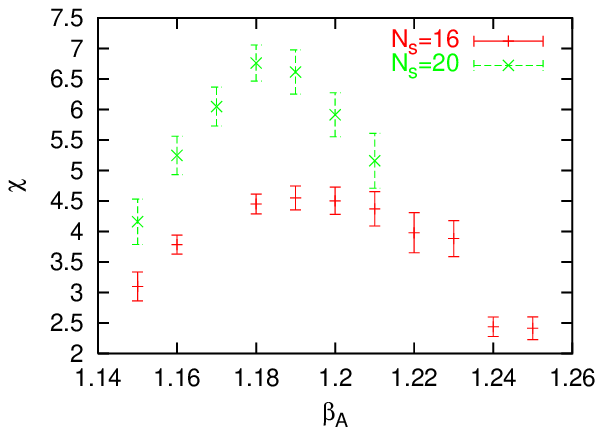}}
\end{center}
\caption{$L_F$ (a) and its susceptibility (b).}
\label{poltr6}
\end{figure}

Fig.~\ref{poltr4} (a, b) show the behaviour of $L_F$ 
and its susceptibility in
the trivial twist sector after the mapping to the positive plaquette model
at $\lambda=1.0$ for $N_{\tau}=4$. In Fig.~\ref{poltr44} (a, b) we perform a FFS 
analysis
with the Ising 3-d critical exponents and our best estimate $\beta_A^c=0.98$,
which agrees with the result in Ref.~\cite{Barresi:2004qa} obtained
through $\rho$.
This is confirmed in Fig.~\ref{poltr6} (a, b), which show $L_F$ and $\chi$ at 
$\lambda=1.0$ for $N_{\tau}=6$. Again, our results are in agreement with the 
estimate $\beta_A^c=1.19$ from $\rho$ in Ref.~\cite{Barresi:2004qa}.

\begin{figure}[htb]
\begin{center}
\subfigure[]{
\includegraphics[angle=0,width=0.45\textwidth]{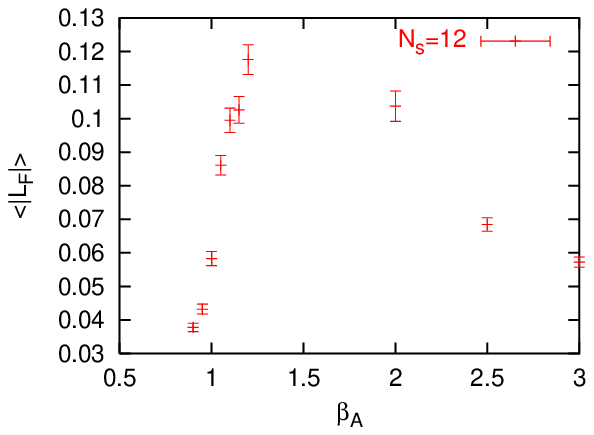}}
\subfigure[]{
\includegraphics[angle=0,width=0.45\textwidth]{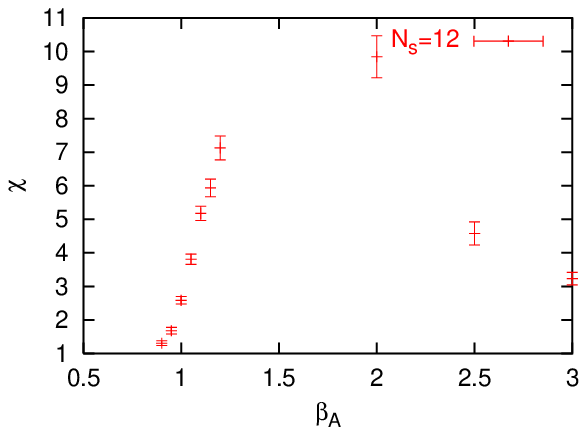}}
\end{center}
\caption{$L_F$ (a) and its susceptibility (b) in the non-trivial twist sector.}
\label{polntr4}
\end{figure}

\begin{figure}[htb]
\begin{center}
\includegraphics[angle=0,width=0.9\textwidth]{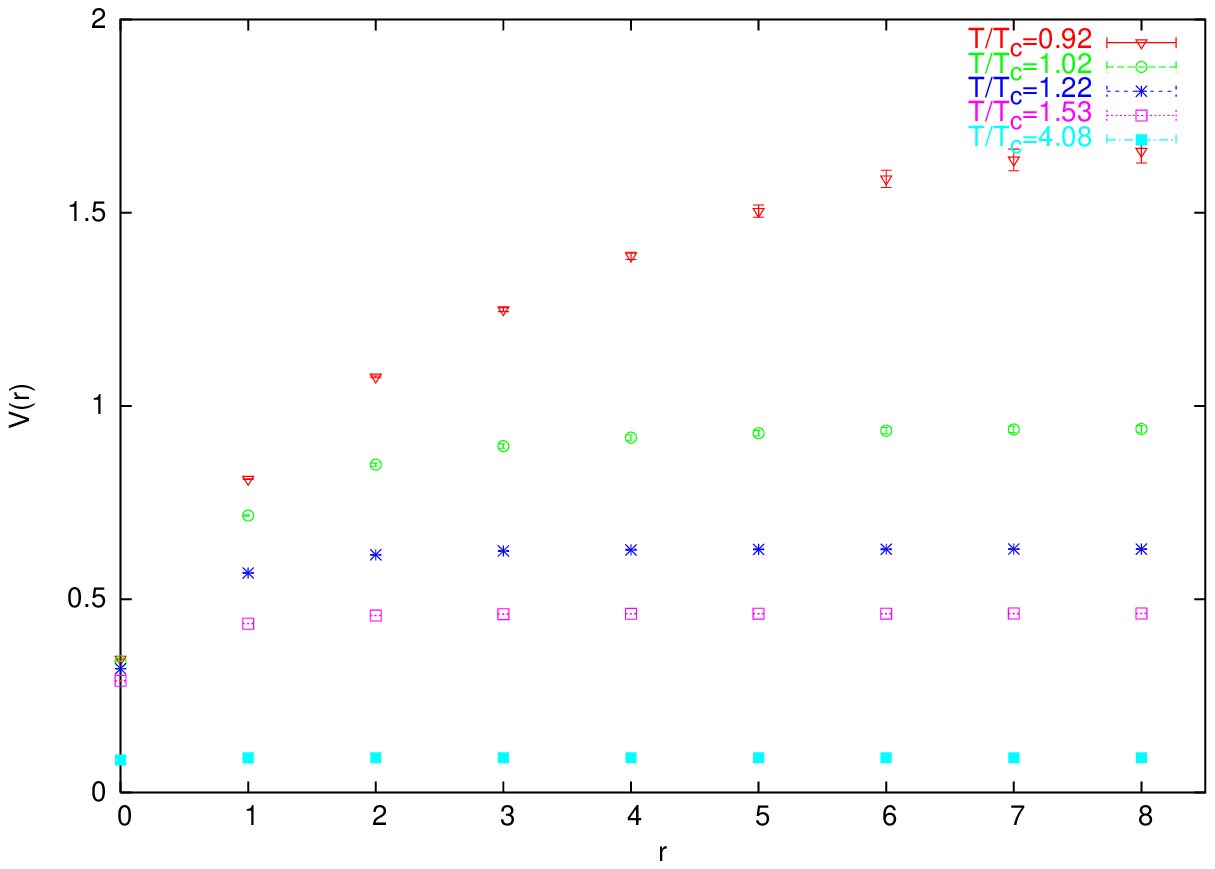}
\end{center}
\caption{$q-\bar{q}$ potential from $L_F$ correlators at trivial twist.}
\label{pot}
\end{figure}

\begin{figure}[htb]
\begin{center}
\includegraphics[angle=0,width=0.9\textwidth]{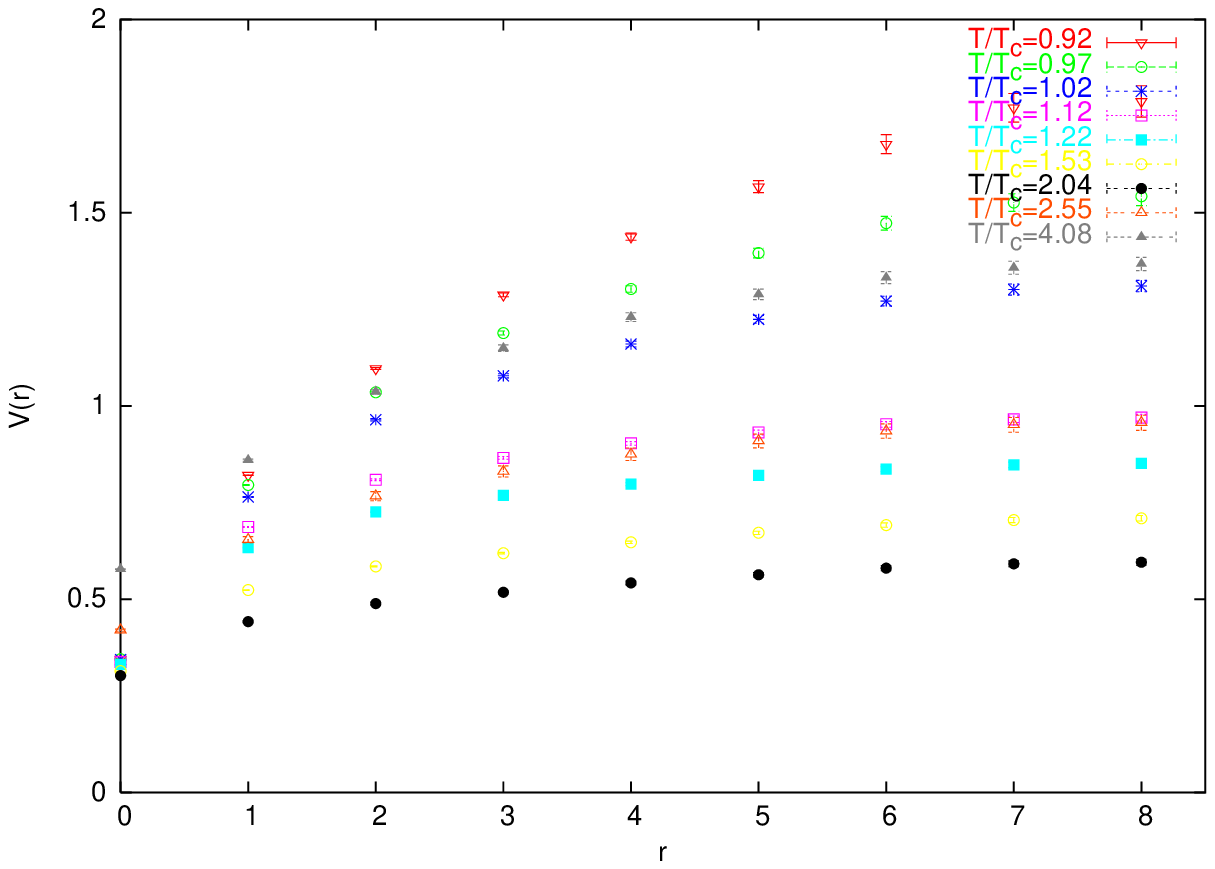}
\end{center}
\caption{$q-\bar{q}$ potential from $L_F$ correlators at non trivial twist.}
\label{pot1}
\end{figure}

\begin{figure}[htb]
\begin{center}
\includegraphics[angle=0,width=0.9\textwidth]{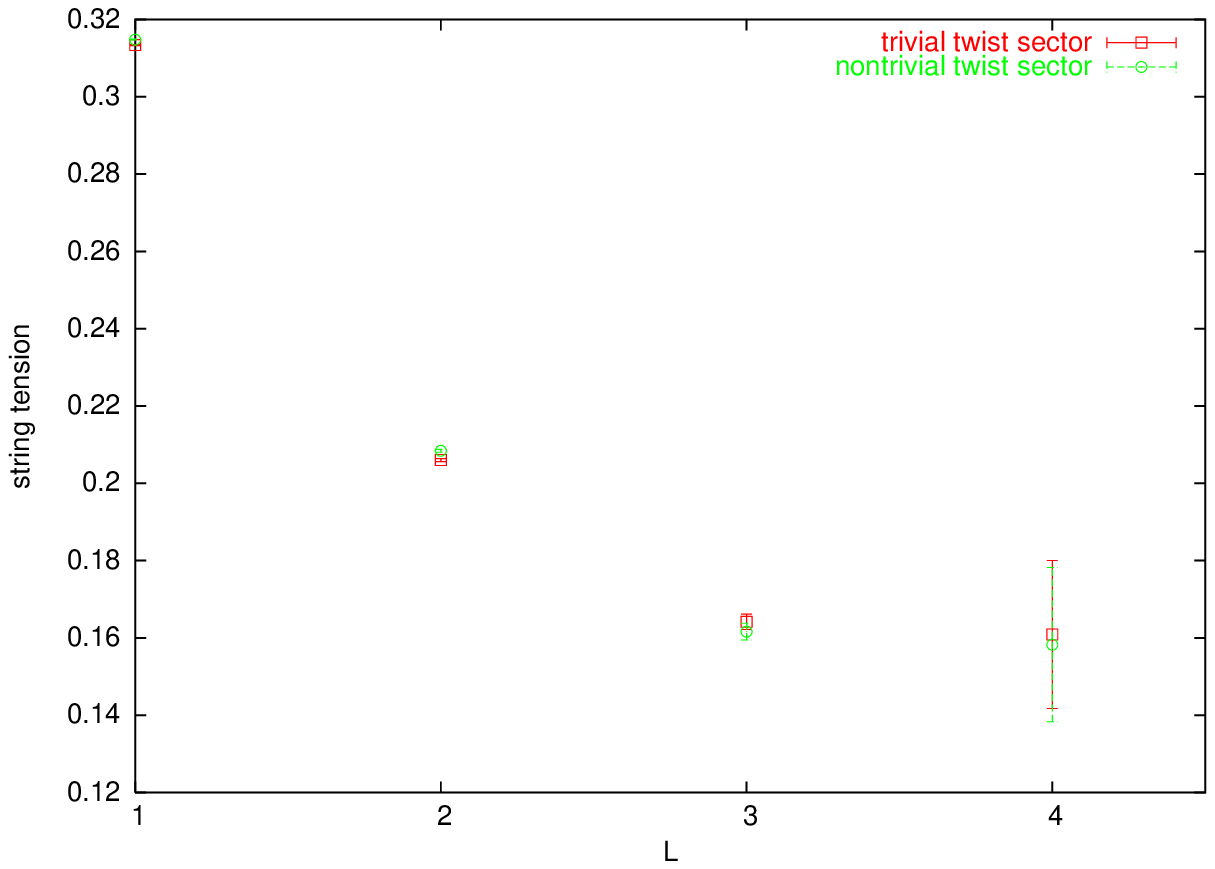}
\end{center}
\caption{String tension estimate for $N_s=N_\tau=16$, $\beta_A=0.98$.}
\label{string}
\end{figure}

The non trivial twist sector offers more material for discussions. 
Fig.~\ref{polntr4} (a,b) show $L_F$ and $\chi$ at $\lambda=1.0$ for 
$N_{\tau}=4$. $L_F$ rises at first above the ``transition'' (with $L_A$ also 
positive) to then tend to
zero in the high $\beta_A$ limit, as expected from the known behaviour
$L_A \to -1/3$. Such behaviour is not that of a standard deconfining theory, 
due to the non trivial background introduced by the non trivial twist. The
determination of $\beta_A^c$ is more difficult, our best estimate remaining
thus that from $\rho$ obtained in Sec.~\ref{sec:pisa}. In light of such problems
and given the worse 
signal to noise ratio at non trivial twist the comparison of Fig.~\ref{nt4}(b)
with the scaling figures for $\rho$ in 
Ref.~\cite{Barresi:2004qa} is not bad. Compare also such result 
with the indubitably better scaling obtained here in
Fig.~\ref{poltr44} to have a measure of the intrinsic difficulties 
in performing precision measurements with $\rho$. 
Figs.~\ref{pot},~\ref{pot1} illustrate perhaps better the situation. There we have 
calculated the interquark potential from $L_F$ 
correlators at $\lambda=1.0$, $N_{\tau}=4$, $N_s=16$ for both twist sectors at
various temperatures as a function of the distance. We have chosen to keep lattice 
units throughout since contrary to the fundamental
case no result exists for the non perturbative scaling of adjoint
models, e.g. through precision measurements of the string tension. 
Below $T_c$ (upper curve in both Figures) both trivial and non trivial twist 
show still a (slowly) linearly growing potential . The growth of $V(r)$ is 
moderate as expected, since the
string tension should get dampened like $\sigma(T) = \sigma \sqrt{1-T^2/T_c^2}$
when approaching a 2$^{\rm nd}$ order transition, i.e.
roughly by a factor 60$\%$ at the value $T= 0.92 T_c$ chosen. 
Above $T_c$ the situation changes. In the trivial twist case the long range 
interactions die very fast above $T_c$, having disappeared at 
$T \simeq 2 T_c$, which corresponds to a lattice spacing roughly the half of that 
at $T_c$.
In the non trivial sector however they persist quite consistently,
so that it is not clear whether one can speak of deconfinement in the
common understanding. The long range interactions are minimal (in lattice units)
at $T \simeq 2 T_c$,
roughly coinciding with the peaks in $L_F$ and its susceptibility, to rise then
again: at $T \simeq 4 T_c$ (which corresponds to 25$\%$ of the $T_c$ lattice spacing) 
they have a similar strength in lattice units as
at $T \simeq T_c$. Of course finite volume effects will start to be considerable at
such relatively high $\beta_A$ values, so that a full analysis would imply going at 
much bigger $N_s$. The difference in qualitative behaviour between Fig.~\ref{pot}
and \ref{pot1} remains anyway striking.
To conclude, Fig.~\ref{string} shows (again in lattice units) the fundamental
string tension estimates in both twist sectors from (fundamental) Creutz ratios 
at $T=0$ ($16^4$), 
$\lambda=1$ and $\beta_A=0.98$. Comparing with the value $a^2 \sigma \simeq 0.140$ 
obtained in the fundamental case at $T=0$ for $\beta \simeq 2.29$, i.e. the coupling
corresponding to $T_c$ for $N_{\tau}=4$, one has the impression that our
positive plaquette model representation of the adjoint theory has no dramatic
scaling discrepancies with the standard fundamental theory.
It would be interesting to recheck these non-trivial twist results either with 
the standard Wilson action or with a genuine positive plaquette model, of
course with twisted boundary condition in both cases. Also the alternative
definition of $L_F$ given in Ref.~ \cite{deForcrand:2001nd} would be worth to
explore. To our knowledge, although twisted b.c. have been used in
the literature, no comparable result to those given here is available.

Fixing a non trivial background through twist sectors leads then to
theories which are not equivalent to the standard one. While
having a similar dynamics below $T_c$ they will
not show the standard deconfining behaviour at high temperature. 
As discussed in section \ref{sec4} there is of course a problem in the 
interpretation of the non trivial
twist sector: even though as we have shown in the ``confined'' phase one can 
formally define an interquark potential and through it a fundamental string 
tension, there is actually no way to couple the gauge fields to 
fermions at finite temperature with twisted b.c. The twisted
adjoint theory, contrary to the untwisted one, cannot therefore be considered 
equivalent to a standard fundamental Yang-Mills theory. $L_F$ and its
correlators cannot therefore be related to the static potential of quarks.
One should therefore not read much in their unusual behaviour. 

\section{Conclusions}

In this paper we have studied the pure adjoint $SU(2)$ theory in the
trivial and non-trivial twist sectors. We have been able to establish the 
properties of the different phases analyzing the Pisa disorder operator;
through an explicit kinematic mapping
to different positive plaquettes models with corresponding boundary 
conditions, we have also studied the behaviour of $L_F$. 
While the adjoint theory in the trivial twist sector can 
be considered equivalent to the standard $SU(2)$ gauge theory, the twisted
theories show a quite different behaviour, in particular exhibiting no 
real deconfinement above $T_c$. In light of this, we are led to conclude that only
the trivial twist sector and the full ergodic theory, i.e. summing over
all twist sectors, can be considered good discretizations of Yang-Mills 
theories: the former gives the quenched theory, the second the full
pure gauge case. 
However, many of our results here obtained for the fixed twisted case, through 
the considerations in Sec.~\ref{sec3}, can be used to establish the properties of the 
ergodic theory in regimes where sampling the full partition function would be
very expensive in computer time. 

\section{Acknowledgements}
We thank A. Di Giacomo, L. Del Debbio, M. M\"uller-Preussker, M. D'Elia 
and P. de Forcrand for valuable comments and discussions. 
A special thank goes to Oliver Jahn for encouraging us to find the explicit 
mappings to positive plaquette models.

\bibliographystyle{elsart}
\bibliography{bib}

\begin{thebibliography}{10}
\expandafter\ifx\csname url\endcsname\relax
  \def\url#1{\texttt{#1}}\fi
\expandafter\ifx\csname urlprefix\endcsname\relax\def\urlprefix{URL }\fi

\bibitem{McLerran:1981pk}
L.~D. McLerran, B.~Svetitsky, Phys. Lett. B98 (1981) 195.

\bibitem{Kuti:1981gh}
J.~Kuti, J.~Polonyi, K.~Szlachanyi, Phys. Lett. B98 (1981) 199.

\bibitem{Polyakov:1978vu}
A.~M. Polyakov, Phys. Lett. B72 (1978) 477.

\bibitem{Susskind:1979up}
L.~Susskind, Phys. Rev. D20 (1979) 2610.

\bibitem{Svetitsky:1982gs}
B.~Svetitsky, L.~G. Yaffe, Nucl. Phys. B210 (1982) 423.

\bibitem{Smilga:1993vb}
A.~V. Smilga, Ann. Phys. 234 (1994) 1.

\bibitem{Bhanot:1981eb}
G.~Bhanot, M.~Creutz, Phys. Rev. D24 (1981) 3212.

\bibitem{Greensite:1981hw}
J.~Greensite, B.~Lautrup, Phys. Rev. Lett. 47 (1981) 9.

\bibitem{Halliday:1981te}
I.~G. Halliday, A.~Schwimmer, Phys. Lett. B101 (1981) 327.

\bibitem{Halliday:1981tm}
I.~G. Halliday, A.~Schwimmer, Phys. Lett. B102 (1981) 337.

\bibitem{Caneschi:1982ik}
L.~Caneschi, I.~G. Halliday, A.~Schwimmer, Nucl.Phys.B200 (1982) 409.

\bibitem{'tHooft:1979uj}
G.~'t~Hooft, Nucl. Phys. B153 (1979) 141.

\bibitem{Mack:1979gb}
G.~Mack, V.~B. Petkova, Zeit. Phys. C12 (1982) 177.

\bibitem{Tomboulis:1980vt}
E.~Tomboulis, Phys. Rev. D23 (1981) 2371.

\bibitem{deForcrand:2002vs}
P.~de~Forcrand, O.~Jahn, Nucl. Phys. B651 (2003) 125.

\bibitem{Cheluvaraja:1996zn}
S.~Cheluvaraja, H.~S. Sharathchandra, hep-lat/9611001.

\bibitem{Datta:1996pi}
S.~Datta, R.~V. Gavai, Phys. Lett. B392 (1997) 172.

\bibitem{Datta:1997nv}
S.~Datta, R.~V. Gavai, Phys. Rev. D57 (1998) 6618.

\bibitem{Datta:1999ep}
S.~Datta, R.~V. Gavai, Phys. Rev. D62 (2000) 054512.

\bibitem{Datta:1999np}
S.~Datta, R.~V. Gavai, Phys. Rev. D60 (1999) 034505.

\bibitem{Burgio:2005xe}
G.~Burgio et al. , PoS LAT2005 (2006) 288.

\bibitem{Burgio:2006dc}
G.~Burgio et al., Phys. Rev. D74 (2006) 071502.

\bibitem{Burgio:2006xj}
G.~Burgio et al. , hep-lat/0610097.

\bibitem{Barresi:2001dt}
A.~Barresi, G.~Burgio, M.~Muller-Preussker, Nucl. Phys. Proc. Suppl. 106 (2002) 495.

\bibitem{Barresi:2002un}
A.~Barresi, G.~Burgio, M.~Muller-Preussker, Nucl. Phys. Proc. Suppl. 119 (2003) 571.

\bibitem{Barresi:2003jq}
A.~Barresi, G.~Burgio, M.~Muller-Preussker, Phys. Rev. D69 (2004) 094503.

\bibitem{Barresi:2003yb}
A.~Barresi, G.~Burgio, M.~Muller-Preussker, Color con\-fi\-ne\-ment-Wako 2003 (2004) 82.

\bibitem{Barresi:2004qa}
A.~Barresi, G.~Burgio, M.~D'Elia, M.~Mueller-Preussker, Phys. Lett. B599 (2004) 278.

\bibitem{Barresi:2004gk}
A.~Barresi, G.~Burgio, M.~Muller-Preussker, Nucl. Phys. Proc. Suppl. 129 (2004) 695.

\bibitem{Fingberg:1995ut}
J.~Fingberg, U.~M. Heller, V.~K. Mitrjushkin, Nucl. Phys. B435 (1995) 311.

\bibitem{DiGiacomo:1997sm}
A.~Di~Giacomo, G.~Paffuti, Phys. Rev. D56 (1997) 6816.

\bibitem{Nambu:1974zg}
Y.~Nambu, Phys. Rev. D10 (1974) 4262.

\bibitem{Mandelstam:1975hb}
S.~Mandelstam, Phys. Rev. D11 (1975) 3026.

\bibitem{'tHooft:1974qc}
G.~'t~Hooft, Nucl. Phys. B79 (1974) 276.

\bibitem{Carmona:2001ja}
J.~M. Carmona, M.~D'Elia, A.~Di~Giacomo, B.~Lucini, G.~Paffuti, Phys. Rev. D64 (2001) 114507.

\bibitem{deForcrand:2001nd}
P.~de~Forcrand, L.~von Smekal, Phys. Rev. D66 (2002) 011504.

\bibitem{Kovacs:2000sy}
T.~G. Kovacs, E.~T. Tomboulis, Phys. Rev. Lett. 85 (2000) 704.

\bibitem{deForcrand:2000fi}
P.~de~Forcrand, M.~D'Elia, M.~Pepe, Phys. Rev. Lett. 86 (2001) 1438.

\bibitem{deForcrand:2005pb}
P.~de~Forcrand, D.~Noth, Phys. Rev. D72 (2005) 114501.

\end{thebibliography}
\end{document}